\documentclass[aps,prl,twocolumn,superscriptaddress]{revtex4-2}

\usepackage{amsthm}
\usepackage{amsmath,bm}
\usepackage{amssymb}
\usepackage{amsfonts}
\usepackage{graphicx}
\usepackage{txfonts}
\usepackage{xcolor}
\usepackage{float}

\usepackage{quantum}

\usepackage[colorlinks=true,linkcolor=blue,citecolor=blue,urlcolor=blue]{hyperref}

\begin{document}

\title{Metrological Sensitivity beyond Gaussian Limits with Cubic Phase States}

\author{Jiajie Guo}
\affiliation{State Key Laboratory for Mesoscopic Physics, School of Physics, Frontiers Science Center for Nano-optoelectronics, $\&$ Collaborative
Innovation Center of Quantum Matter, Peking University, Beijing 100871, China}

\author{Shuheng Liu}
\affiliation{State Key Laboratory for Mesoscopic Physics, School of Physics, Frontiers Science Center for Nano-optoelectronics, $\&$ Collaborative
Innovation Center of Quantum Matter, Peking University, Beijing 100871, China}

\author{Boxuan Jing}
\affiliation{State Key Laboratory for Mesoscopic Physics, School of Physics, Frontiers Science Center for Nano-optoelectronics, $\&$ Collaborative
Innovation Center of Quantum Matter, Peking University, Beijing 100871, China}

\author{Qiongyi He}
\email{qiongyihe@pku.edu.cn}
\affiliation{State Key Laboratory for Mesoscopic Physics, School of Physics, Frontiers Science Center for Nano-optoelectronics, $\&$ Collaborative Innovation Center of Quantum Matter, Peking University, Beijing 100871, China}
\affiliation{Collaborative Innovation Center of Extreme Optics, Shanxi University, Taiyuan, Shanxi 030006, China}
\affiliation{Peking University Yangtze Delta Institute of Optoelectronics, Nantong, Jiangsu 226010, China} 
\affiliation{Hefei National Laboratory, Hefei 230088, China}

\author{Manuel Gessner}
\email{manuel.gessner@uv.es}
\affiliation{Instituto de Física Corpuscular (IFIC), CSIC-Universitat de València and Departament de Física Teòrica, UV, C/Dr Moliner 50, E-46100 Burjassot (Valencia), Spain }

\begin{abstract}
Cubic phase states provide the essential non-Gaussian resource for continuous-variable quantum computing. We show that they also offer significant potential for quantum metrology, surpassing the phase-sensing sensitivity of all Gaussian states at equal average photon number. Optimal sensitivity requires only moderate initial squeezing, and the non-Gaussian advantage remains robust against loss and detection noise. We identify optimal measurement strategies and show that several experimentally relevant preparation schemes surpass Gaussian limits, in some cases reaching the sensitivity of cubic phase states. Our results establish cubic phase states as a promising resource for quantum-enhanced precision measurements beyond Gaussian limits. 
\end{abstract}

\maketitle

\textit{Introduction.---}Continuous-variable (CV) quantum systems are a central platform for implementations of quantum information science and quantum technologies~\cite{BraunsteinRMP2005,SerafiniBook2017,AdessoOSID2014,WeedbrookRMP2012,WangPR2007}. In this setting, Gaussian states have been studied extensively due to their experimental accessibility and their convenient description in terms of first and second moments~\cite{WeedbrookRMP2012,WangPR2007,AlessandroArXiv2005,JonatanArXiv2022}. Non-Gaussian states, however, can outperform Gaussian states in several protocols~\cite{UlrikNp2015,MattiaPRXQ2021,FabianNC2019}, such as quantum error correction~\cite{CampagneNature2020}, secure quantum commutation~\cite{Jaehaknpj2019}, quantum key distribution~\cite{GuoPRA2019} and quantum computing~\cite{MariPRL2012,TakagiPRA2018,AlbarelliPRA2018,ChabaudPRL2023}.

In quantum metrology, non-Gaussian states play a particularly important role in multiparticle atomic systems, where the finite surface of the Bloch sphere imposes an upper bound on the degree of spin squeezing for a fixed particle number~\cite{PezzeRMP2018}. In this regime, non-Gaussian spin states such as GHZ states and over-squeezed states are essential for attaining quantum advantages beyond those of approximately Gaussian spin-squeezed states~\cite{HelmutScience2014,PezzeRMP2018,KaiPRL2022,SimoneNP2022,ManuelPRL2019,BaamaraPRL2021,BaamaraCRP2022,GuoPRA2023,GuoArXiv2025}. In contrast, CV platforms such as optical fields and mechanical oscillators have an infinite phase space, implying that there is no fundamental upper bound on the amount of squeezing that can be generated. This naturally kept the focus on Gaussian strategies in CV quantum metrology. 

This emphasis on Gaussian states is further supported by two considerations: First, for displacement estimation, Gaussian probe states are known to be optimal~\cite{MatteoRPP2025}. Second, in rotation or phase estimation, squeezed vacuum states--the optimal Gaussian probes~\cite{MonrasPRA2006,MonrasArXiv2013,PinelPRA2013,TeruoNJP2019}--outperform a broad class of non-Gaussian states investigated so far, including cat states, compass states~\cite{ZurekNature2001}, and balanced superpositions of Fock states~\cite{MatteoRPP2025}. Since no state-independent upper bound on phase sensitivity exists at fixed mean population $n$, one may in principle design probe states with arbitrary large number fluctuations~\cite{ShapiroPRL1989,RivasNJP2012} that would achieve correspondingly large sensitivity. However, such unbounded sensitivities--particularly those that surpass the $n^2$-scaling of the optimal Gaussian strategy--have been shown to be unattainable in practice when concrete readout and estimation schemes are considered~\cite{BraunsteinPRL1992,LanePRA1993,LacaPRA2013,GiovannettiPRL2012,HyllusPRL2010}. Thus, to date, no viable alternatives have been shown to surpass the sensitivity scaling of squeezed-vacuum states.

\begin{figure}[t]
    \centering
    \includegraphics[width=.48\textwidth]{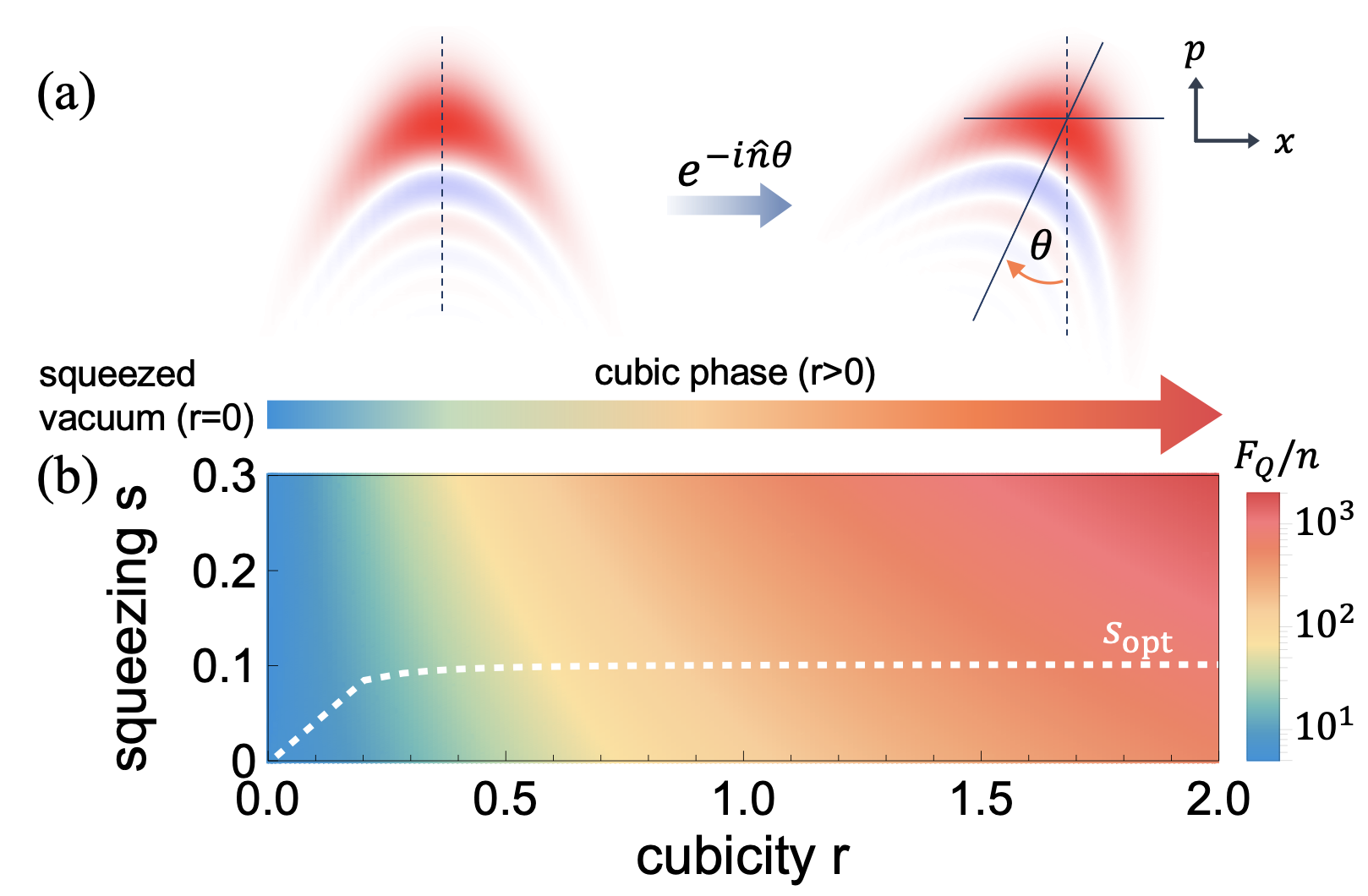}
    \caption{ \textbf{Rotation sensing with a cubic phase state.} (a) Wigner function of a cubic phase state under the rotation $e^{-i \hat{n} \theta}$, corresponding to a phase-space rotation around the origin. (b) Sensitivity $F_Q(r,s)/n$ as a function of squeezing $s$ and cubicity $r$. The metrological potential increases rapidly with $r>0$, with cubic phase states significantly outperforming squeezed vacuum states ($r=0$). The white dashed line indicates the optimal squeezing $s_{\text{opt}}$.}
    \label{Fig1}
\end{figure}

The situation differs in CV quantum computing, where purely Gaussian operations are known to be classically simulable~\cite{MariPRL2012,TakagiPRA2018,AlbarelliPRA2018,ChabaudPRL2023}. To overcome this limitation, cubic phase states were introduced as the essential non-Gaussian resources enabling universal CV quantum computation~\cite{GottesmanPRA2001,GuPRA2009,MenicucciPRL2006,MenicucciPRL2014,BudingerPRR2024}. Although their experimental realization is demanding--owing to the lack of a natural physical process that directly implements the cubic phase gate $e^{i r\hat{x}^3}$--substantial progress has been made in recent years. In addition to the deterministic generation of exact cubic phase states in superconducting microwave circuits~\cite{ErikssonNC2024}, several schemes for the approximate realization of the cubic gate have been developed~\cite{MarshallPRA2015,MarekPRA2011,MiyataPRA2016,JingarXiv2025,YanagimotoPRL2020,ZhengPRXQ2021}. These rely on non-Gaussian measurements or higher-order interactions and have enabled the experimental preparation of approximate cubic phase states in optical platforms~\cite{YukawaPRA2013}. Despite their central role in CV quantum computation, the metrological potential of cubic phase states has so far remained unexplored.

In this work, we show that cubic phase states represent highly sensitivite probe states for CV quantum phase estimation. Specifically, they attain a quantum Fisher information scaling as $F_Q\sim (128/3)n^2\approx 42.7n^2$, exceeding by more than a factor of five the maximum sensitivity achievable with Gaussian states, which are limited to $F_Q\sim 8n^2$. To assess practical implementability, we employ nonlinear squeezing coefficients based on simple moment-based estimators that use the average values of optimally chosen nonlinear observables. We show that fourth-order moments are optimal, as they saturate the quantum Cramér–Rao bound, and we find that this approach remains reasonably robust to loss and detection noise. We further show that existing protocols for preparing approximate cubic phase states can already deliver notable sensitivity gains. Together, these results identify cubic phase states as promising non-Gaussian resources for CV phase sensing and highlight feasible pathways toward sensitivities beyond the Gaussian regime in current experimental platforms.

\textit{Rotation sensing with a cubic phase state.---}A cubic phase state is prepared by applying the cubic gate $\hat{C}(r)=e^{i r \hat{x}^3}$ to a momentum-squeezed vacuum state, yielding
\begin{align}\label{eq:cubicphase}
|\psi (r,s) \rangle &= e^{i r \hat{x}^3 } e^{ -s (\hat{a}^2 -\hat{a}^{\dagger 2})/2 } |0\rangle,
\end{align}
where $s$ denotes the squeezing strength and the cubicity $r$ introduces a nonlinear deformation that shears and bends the state towards the $-p$ direction, producing a strongly non-Gaussian state. 

In a rotation-sensing protocol, the objective is to estimate a phase shift $\theta$ imprinted on a probe state $\rho$ by the unitary operation $e^{ -i\theta \hat{n}}$, with $\hat{n}=\hat{a}^\dagger \hat{a}$ the number operator. As illustrated in Fig.~\ref{Fig1} (a), this operation corresponds to a rotation of the state in phase space around the origin, yielding $\rho(\theta) = e^{ -i\theta \hat{n}} \rho e^{ i\theta \hat{n}}$. The ultimate quantum precision limit achievable by any unbiased estimator $\theta_{\mathrm{est}}$ is set by the quantum Cram$\acute{\text{e}}$r-Rao bound $\Delta \theta_{\mathrm{est}} \geq \Delta \theta_{QCR} \equiv 1/\sqrt{\mu F_Q[\rho, \hat{n}]}$~\cite{BraunsteinPRL1994}, where $F_Q[\rho,\hat{n}]$ is the quantum Fisher information (QFI) that quantifies the sensitivity of $\rho$ to perturbations generated by $\hat{n}$, and $\mu$ is the number of independent measurements.

For pure states, the QFI is determined by the variance of the generator, $F_Q [\psi, \hat{n}] = 4\text{Var}[\psi, \hat{n}]$, and for the state $|\psi (r,s) \rangle$ in Eq.~\eqref{eq:cubicphase}, its analytical expression $F_Q(r,s)=F_Q [\psi(r,s), \hat{n}]$ in terms of $s$ and $r$ is 
\begin{align}\label{eq:QFIrs}
F_Q(r,s)&= 486 e^{8s} r^4+\left(\frac{9}{2} e^{2s} +27 e^{6s} \right) r^2+ \cosh{(4s)}  -1.
\end{align}
To evaluate the sensitivity at fixed quantum resources, we express all quantities in terms of the average particle number
\begin{align}\label{eq:population}
n = \langle \hat{n} \rangle_{\ket{\psi(r,s)}} = \sinh^2{(s)} + \frac{27}{8} e^{4s} r^2,
\end{align}
which shows that squeezing $s$ increases the population exponentially, while the contribution from the cubicity $r$ grows only quadratically.

Using Eq.~\eqref{eq:population}, the QFI~\eqref{eq:QFIrs} can be rewritten as a quadratic polynomial in the population $n$
\begin{align}\label{eq:QFIns}
    F_Q (n,s) = c_2 n^2 + c_1(s)n + c_0(s),
\end{align}
where the leading coefficient $c_2=128/3$ is independent of $s$. The remaining coefficients are $c_1(s)=a(s)-2c_2\sinh^2(s)$, and $c_0(s)=[8 -a(s) + (8 + c_2) \sinh^2(s)] \sinh^2(s)$, with $a(s)=4/3 e^{-2  s} + 8  e^{2  s}$. Note that Eq.~\eqref{eq:population} implies $n \geq \sinh^2{(s)}$; consequently $n=0$ entails $s=0$.

In the large-$n$ limit, the QFI approaches
\begin{align}
F_Q \sim \frac{128}{3} n^2 \approx 42.7 n^2.
\end{align}
which exceeds the optimal Gaussian scaling achieved by squeezed vacuum states ($F_Q \sim 8n^2$) and also surpasses several non-Gaussian strategies, including using large ($|\alpha| \gtrsim 2$) cat states ($F_Q\sim 4n$), compass states ($F_Q \sim 4n$), and balanced Fock superposition states ($ F_Q \sim 4n^2 $)~\cite{MatteoRPP2025}. 

The origin of this enhancement can be understood from the state's Wigner function, shown in Fig.~\ref{Fig1}(a). The strongly sheared distribution features fine interference fringes that are nearly orthogonal to the rotation axis while remaining locally parallel, so that small rotations shift the fringes almost directly on top of one another, causing pronounced state variations and hence high phase sensitivity. Figure~\ref{Fig1}(b) shows that the cubicity $r$ is the dominant parameter for enhancing the sensitivity $F_Q(r,s)/n$. At $r=0$, the states reduce to squeezed vacuum states with comparatively modest sensitivity---although, as mentioned above, still outperforming many non-Gaussian benchmarks. For $r>0$, the cubic phase gate introduces non-Gaussian features that lead to a rapid enhancement of the sensitivity $F_Q/n$ with growing $r$.

\begin{figure}[t]
    \centering
    \includegraphics[width=90mm]{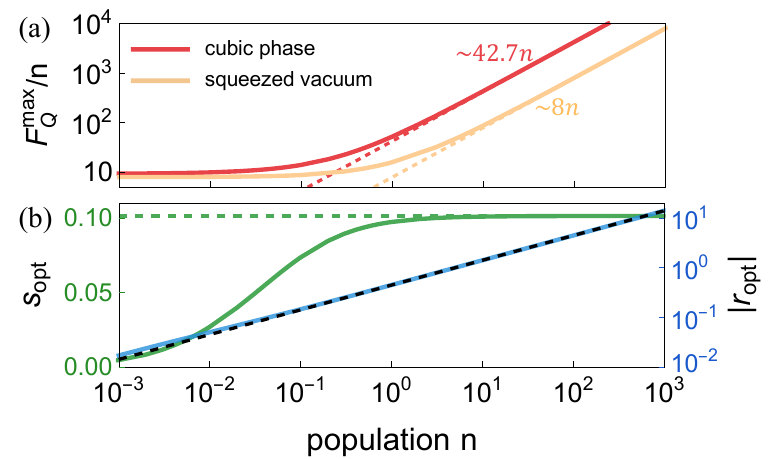}
    \caption{ \textbf{ Metrological scaling laws for cubic phase states.} (a) Maximum sensitivity gain for cubic phase states (red solid line) compared with squeezed vacuum states (yellow solid line). In the limit of large population $n$, cubic phase states scale as $F_Q/n \sim \frac{128}{3}n \approx 42.7n $ (red dashed line), clearly exceeding the squeezed-vacuum scaling $F_Q/n \sim 8n $ (yellow dashed line). (b) Optimal squeezing $s_{\text{opt}}$ (green solid line) and cubicity $|r_{\text{opt}}|$ (blue solid line) that achieve the maximum $F^{\max}_Q/n$. Asymptotically, they approach $s_{\text{opt}}^{n\rightarrow \infty} = \frac{1}{2} \log{\left( \sqrt{6}/2 \right)} \approx 0.1014 $ (green dashed line), corresponding to only a moderate level of $\sim 0.88\,\mathrm{dB}$ of squeezing, and $|r_{\text{opt}}^{n\rightarrow \infty }| = \frac{4\sqrt{n}}{9}$ (black dashed line).}
    \label{Fig2}
\end{figure}

For arbitrary $n$, the optimal squeezing $s_{\text{opt}}$ that maximizes the sensitivity at fixed $n$ is obtained from Eq.~\eqref{eq:QFIns}, $F_Q^{\text{max}}(n)=F_Q(n,s_{\text{opt}})=\max_s F_Q(n,s)$. Once $s=s_{\text{opt}}$ is fixed, the corresponding cubicity $|r_{\text{opt}}|$ is uniquely determined by the chosen population $n$ through Eq.~(\ref{eq:population}). In the large-$n$ limit, these parameters approach
\begin{align}
s^{n\rightarrow \infty}_{\text{opt}} &= \frac{1}{2} \log{\left( \frac{\sqrt{6}}{2} \right)} \approx 0.1014, \\
| r^{n\rightarrow \infty}_{\text{opt}} | &= \frac{4\sqrt{n}}{9}.
\end{align}
This shows that preparing highly sensitive cubic phase states requires only moderate squeezing. For finite $n$, the optimal squeezing is upper bounded by its asymptotic value, $s_{\text{opt}}\leq s^{n\rightarrow \infty}_{\text{opt}}$, corresponding to an input state with $0.88\,\mathrm{dB}$ of squeezing. The dominant impact on the state's sensitivity stems from the cubicity $r$ which only consumes population at a rate of $\sim n^{1/2}$. The derivations of the QFI and optimal choice of parameters are provided in SM~\cite{SM} Sec. I.

Figure~\ref{Fig2}~(a) compares the maximal sensitivity $F^{\text{max}}_Q(n)/n$ of cubic phase states with that of squeezed vacuum states, showing a clear advantage for all $n$. Figure~\ref{Fig2}~(b) shows the corresponding optimal parameters: the optimal squeezing $s_{\text{opt}}$ increases with population and rapidly approaches its asymptotic value (green dashed) around $n\approx 1$, whereas the optimal cubicity $|r_{\text{opt}}|$ closely follows its asymptotic $\sim n^{1/2}$ scaling (black dashed) over the entire range.

\textit{Practical non-Gaussian sensitivity advantage with nonlinear squeezing parameters.---}Extracting the full sensitivity of non-Gaussian probe states requires access to higher-order moments, since linear quadrature measurements only capture their Gaussian features. To develop a practical strategy that exploits the properties of cubic phase states, we now analyze a family of lower bounds on the achievable sensitivity using simple moment-based estimators built from any accessible observable $\hat{M}$. By considering $\hat{M}$ to be an optimal linear combination of quadrature moments $\mathbf{M}_{(k)}$ up to order $k$, we obtain the nonlinear squeezing coefficient~\cite{ManuelPRL2019},
\begin{align}
\xi_{(k)}^{-2} := \max_{ \hat{M} \in \mathbf{M}_{(k)}}\frac{ | \langle [\hat{n}, \hat{M} ]\rangle_{\rho} |^2 }{ n\text{Var}[\rho,\hat{M}] } ,
\end{align}
where the optimal observable is identified analytically. The resulting coefficients form a hierarchy of lower bounds on the state's full sensitivity $\xi_{(1)}^{-2} \leq \xi_{(2)}^{-2} \leq \cdots \leq F_Q/n$ with the linear squeezing coefficient $\xi_{(1)}^{-2}$ at the lowest order.

\begin{figure}[t]
    \centering
    \includegraphics[width=85mm]{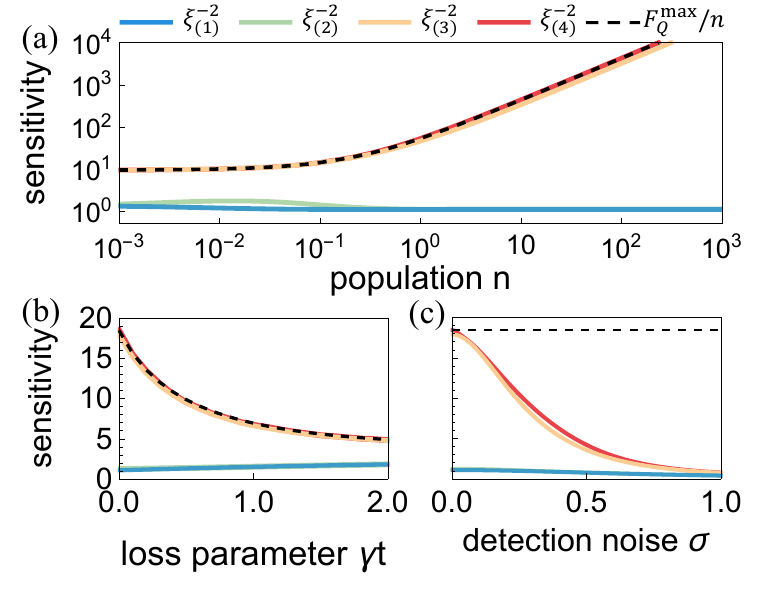}
    \caption{\textbf{Non-Gaussian sensitivity revealed by nonlinear squeezing coefficients.} (a) The squeezing coefficients $\xi_{(i)}^{-2}$ obtained from nonlinear measurements of order up to $i=\{1,2,3,4\}$ for the cubic phase state with optimal parameters $(s_{\text{opt}}, |r_{\text{opt}}|)$ as a function of population $n$. In the hierarchy $\xi_{(1)}^{-2} \leq \xi_{(2)}^{-2} \leq \xi_{(3)}^{-2} \leq \xi_{(4)}^{-2} $ the fourth-order squeezing coefficient is found to exactly saturate the ultimate limit, $\xi_{(4)}^{-2} = F^{\text{max}}_Q/n$, while the third order $\xi_{(3)}^{-2}$ already follows it almost tightly. At $n=0.2$, we present the decay of the sensitivity as a function of (b) $\gamma t$, where $\gamma$ is the loss rate and the time evolution is fixed as $t=1$, and (c) the standard deviation $\sigma$ of Gaussian detection noise.  }
    \label{Fig3}
\end{figure}

We find that the following observable sets suffice to optimize sensitivity up to order $k$: $\mathbf{M}_{(1)} = \{ \hat{x} \} $, $\mathbf{M}_{(2)}=\{ \mathbf{M}_{(1)}, \mathcal{S}(\hat{x}\hat{p}) \}$, $\mathbf{M}_{(3)}=\{ \mathbf{M}_{(2)}, \hat{x}^3, \mathcal{S}(\hat{x}\hat{p}^2) $, $\mathbf{M}_{(4)} = \{ \mathbf{M}_{(3)}, \mathcal{S}\left( \hat{x}^3\hat{p} \right),  \mathcal{S}( \hat{x}\hat{p}^3 ) \}$.  Here $\mathcal{S}(\hat{x}^{k-m} \hat{p}^m ) = \frac{1}{k!} \sum_{ \mathcal{P} \in \mathcal{S}_k} \mathcal{P}\left( \hat{x}^{k-m} \hat{p}^m \right)$, denotes the fully symmetrized operator, with $\mathcal{P}$ running over all permutations of the $k$ entries. Explicit optimal linear combinations at each order are given in the SM~\cite{SM} Sec. II B, and illustrated in Fig.~\ref{Fig3}(a). While first- and second-order nonlinear squeezing coefficients are insufficient to capture the highly sensitive non-Gaussian structure of cubic phase states, the  third- and the fourth-order coefficients take the form
\begin{align}
\xi_{(3)}^{-2} (n,s) &= 32 n + \mathcal{C}_1(s) + \mathcal{C}_2(s) n^{-1},\\
\xi_{(4)}^{-2} (n,s) &= F_Q(n,s)/n,
\end{align}
with $\mathcal{C}_i(s)$ denoting $s$-dependent constants that can be found in Ref.~\cite{SM} Sec. II B.  The third-order scaling $\xi_{(3)}^{-2} \sim 32 n$ already exceeds the optimal Gaussian limit ($F_Q/n \sim 8n$). Remarkably, the fourth-order coefficient $\xi_{(4)}^{-2}$ saturates the quantum Cram$\acute{\text{e}}$r-Rao bound, demonstrating that an optimal measurement achieving the QFI is contained in $\mathbf{M}_{(4)}$. We further show that the corresponding estimators are locally unbiased and analyze their detection statistics, clarifying why this protocol avoids the shortcomings that render sub-Heisenberg strategies unviable (see SM~\cite{SM} Sec. II C).

\textit{Noise robustness.---}
For continuous variable systems, a primary source of decoherence is energy relaxation. The resulting non-unitary loss channel drives the ideal probe state~(\ref{eq:cubicphase}) into a mixed state, with an evolution governed by a Lindblad master equation with a jump operator $\sqrt{\gamma} \hat{a}$~\cite{SerafiniBook2017,BreuerBook2010}. Since particle loss directly affects the properties of the probe state, the ultimate sensitivity $F_Q/n$ decays with increasing $\gamma t$. Nonetheless, we observe in Fig.~\ref{Fig3}~(b) that both nonlinear squeezing coefficients $\xi_{(3)}^{-2}$ and $\xi_{(4)}^{-2}$ follow $F_Q/n$ tightly throughout the entire loss range. When half of the excitations are lost ($\gamma t\approx 0.69$), the probe still retains about $46\%$ of its lossless sensitivity. This level of robustness is compatible with CV platforms, including superconducting circuits, where excitation losses as low as $0.2\%$ have been reported in Ref.~\cite{ErikssonNC2024}.

Another experimentally relevant imperfection is detection noise, which is particularly relevant for the higher-order moment measurements due to their increased statistical demands. Detection noise does not modify the quantum state and therefore leaves the QFI constant. It only limits the performance of specific measurement strategies, such as those characterized by the nonlinear squeezing coefficients, as is illustrated in Fig.~\ref{Fig3}~(c) for Gaussian  noise with the standard deviation $\sigma$. For noise at the level of half vacuum fluctuation, $\sigma=1/(2\sqrt{2})$, the nonlinear coefficients $\xi_{(3)}^{-2}$ and $\xi_{(4)}^{-2}$ drop to $35.6\%$ and $39.8\%$ of their ideal values, respectively,
yet they retain an advantage over linear strategies up to noise levels as large as $\sigma \approx 1$. A detailed discussion on noise robustness is presented in SM~\cite{SM} Sec. III.

\textit{Protocols for approximate cubic phase states.---}The nonlinear cubic phase gate, $\hat{C}(r)=e^{i r \hat{x}^{3}}$, is challenging to implement experimentally, as it does not arise naturally in most physical systems. A notable exception is provided by superconducting microwave circuits, where suitable combinations of flux and charge drives~\cite{HillmannPRL2020} enable a native realization of cubic interactions, recently used to generate true cubic phase states~\cite{ErikssonNC2024}. In other platforms, several protocols have been developed to effectively approximate this nonlinear operation~\cite{MarekPRA2011,MarshallPRA2015,MiyataPRA2016,JingarXiv2025,YanagimotoPRL2020,ZhengPRXQ2021}, leading to the experimental preparation of approximate cubic phase states in optical settings~\cite{YukawaPRA2013}. In this section, we analyze the metrological performance of such practical schemes.

\begin{figure}[t]
    \centering
    \includegraphics[width=88mm]{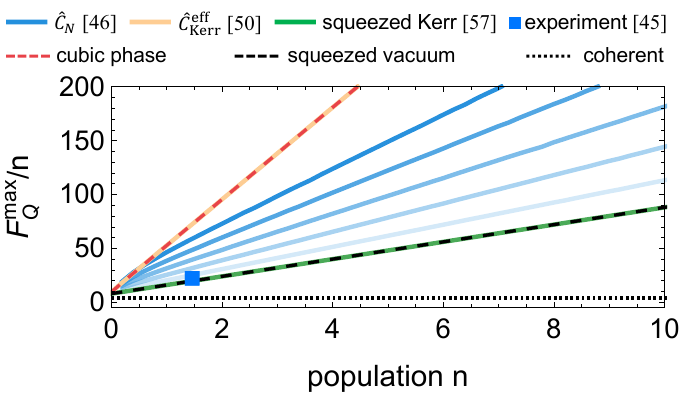}
    \caption{ \textbf{Metrological potential of practical approximate cubic phase states.} The maximum sensitivity gains $F_Q/n$ of practical preparation schemes--sequential photon subtractions~\cite{MarshallPRA2015} (blue solid line, darker tones indicate larger iterations $N=1,\dots,5$), squeezed Kerr dynamics~\cite{GuoPRA2024,VenkatramanPNAS2024} (green solid line), and Kerr interaction combined with Gaussian channels~\cite{YanagimotoPRL2020} (yellow solid line)--are shown as a function of population $n$. These results are compared to the benchmarks provided by ideal cubic phase states ($F_Q/n \sim 42.7 n$, red dashed line), squeezed vacuum states ($F_Q/n \sim 8 n$, black dashed line), and coherent states ($F_Q/n \sim 4$, black dotted line). The blue square indicates the sensitivity inferred from experimental tomographic data of Ref.~\cite{ErikssonNC2024}.  }
    \label{Fig4}
\end{figure}

One approach to preparing cubic phase states is to combine non-Gaussian detection with Gaussian operations. A representative protocol is given in~\cite{MarshallPRA2015}, where the cubic phase gate $e^{i r \hat{x}^{3}}$ is approximated by the decomposition
\begin{align}
\hat{C}_{N} (r) &= \left( 1 + i \frac{r}{N} \hat{x}^3  \right)^N.
\end{align}
Each factor $1 + i \frac{r}{N} \hat{x}^3$ can be implemented using Gaussian operations together with sequential photon subtractions, and $N$ denotes the number of iterations. In the limit of large $N$, this decomposition converges to the ideal cubic phase gate, $\hat{C}_{N \rightarrow \infty} (r) = \hat{C} (r)$. The maximum sensitivity $F_{Q}/n$ of the states $|\psi\rangle= \mathcal{N} \hat{C}_{N}(r) e^{-s\left( -\hat{a}^{\dagger 2} + \hat{a}^2 \right)/2} |0\rangle $ with normalization $\mathcal{N}$ for $N=1,\dots,5$ is shown by the blue lines in Fig.~\ref{Fig4}, where darker tones indicate larger $N$. Already at $N=1$, the sensitivity exceeds that of squeezed vacuum states (black dashed). As $N$ increases, the sensitivity improves and progressively approaches that of the ideal cubic phase states (red dashed).

Another route to generating cubicity is to exploit intrinsic nonlinear interactions. A paradigmatic example is the Kerr interaction $(\sim \hat{a}^{\dagger 2} \hat{a}^2)$, which naturally appears in many CV systems, including in superconducting microwave circuits. However, when applied directly to rotation sensing, the Kerr interaction is ineffective: Since $[ e^{i K \hat{a}^{\dagger 2} \hat{a}^2 }, \hat{n}]=0$, a standalone Kerr evolution does not enhance sensitivity. The Hamiltonian $\hat{H} = \Delta \hat{a}^\dagger \hat{a} + s\left( \hat{a}^{\dagger 2} +\hat{a}^2 \right) -K \hat{a}^{\dagger 2} \hat{a}^2 $ with detuning $\Delta$ is experimentally relevant~\cite{GuoPRA2024,VenkatramanPNAS2024}, but also fails to enhance sensitivity beyond that of squeezing vacuum states (green curve in Fig.~\ref{Fig4}). 

A Kerr-based protocol that effectively approximates a cubic phase gate was proposed in Ref.~\cite{YanagimotoPRL2020}. Embedded between two Gaussian operations, the process described by
\begin{align}
\hat{C}_{\text{Kerr}}^{\text{eff}} = \hat{S}^{\dagger} (\xi) \hat{D}^\dagger (\alpha) e^{-i \hat{H}_{\text{Kerr}} \tau } \hat{D} (\alpha) \hat{S}(\xi),
\end{align}
employs the Kerr-based Hamiltonian $\hat{H}_{\text{Kerr}} = -(K/2) \hat{a}^{\dagger 2} \hat{a}^2 + \Delta \hat{a}^\dagger \hat{a} + \beta ( \hat{a} + \hat{a}^\dagger )$, with a driving amplitude $\beta$, and $\hat{S}(\xi)$ and $\hat{D}(\alpha)$ are squeezing and displacement operators, respectively. By optimizing the parameters of the Kerr Hamiltonian and the Gaussian channels, we find that the resulting states  $ \hat{C}_{\text{Kerr}}^{\text{eff}}(r) e^{-s(\hat{a}^2-\hat{a}^{\dagger 2})/2} |0\rangle $ attain the same maximum sensitivity $F^{\text{max}}_Q/n$ as the ideal cubic phase states (yellow line in Fig.~\ref{Fig4}). 

A trisqueezing interaction
\begin{align}
\hat{U}_{\text{T}}= e^{i \left( t^* \hat{a}^{3} + t\hat{a}^{\dagger 3} \right)}
\end{align}
with the triplicity parameter $t$, has been realized experimentally via three-photon spontaneous parametric down-conversion~\cite{ChangPRX2020}, can also be used to approximate cubic phase states with additional Gaussian channels~\cite{ZhengPRXQ2021}. This interaction was also at the heart of the successful generation of an exact cubic phase state in a superconducting microwave circuit~\cite{ErikssonNC2024}. A maximum-likelihood reconstruction of the density matrix from the reported Wigner tomography data~\cite{ErikssonNC2024} shows that the produced state exceeds the sensitivity of squeezed vacuum states in spite of experimental noise that limits it from reaching the ideal state's full potential (see blue square in Fig.~\ref{Fig4}). The metrological sensitivity of trisqueezed states $\hat{U}_{\text{T}}|0\rangle$ surpasses that of cubic phase states at small populations $n$ (see SM~\cite{SM} Sec. IV C), their intrinsic oscillatory dynamics~\cite{SahelNJP2025} prevents these states from reaching large populations $n$, implying that any scalable family of probe states whose sensitivity grows with $n$ will ultimately outperform them.

\textit{Conclusions.---}We have shown that cubic phase states provide substantial quantum advantages in phase estimation beyond Gaussian limits. Analytical results for the QFI and nonlinear squeezing coefficients demonstrate that their sensitivity is fully captured by the squeezing of fourth-order quadrature observables, which remain reasonably robust against loss and detection noise. We further find that certain experimentally relevant approximate generation schemes can reproduce this advantage to a useful degree in existing CV platforms. Our work advances the fundamental limits of CV phase sensing and establishes that cubic phase gates play a distinctive role not only in CV quantum computation but also in quantum metrology.

\textit{Acknowledgments.---}This work is supported by Quantum Science and Technology-National Science and Technology Major Project (Grant No. 2024ZD0302401 and No. 2021ZD0301500), National Natural Science Foundation of China (No. 12125402, No. 12534016, No. 12505010, No. 12447157, No. 12405005), and Beijing Natural Science Foundation (Grant No. Z240007). J.G. acknowledges Postdoctoral Fellowship Program of CPSF (GZB20240027), and the China Postdoctoral Science Foundation (No. 2024M760072). S.L. acknowledges the China Postdoctoral Science Foundation (No. 2023M740119). M.G. is supported by the project PID2023-152724NA-I00, with funding from MCIU/AEI/10.13039/501100011033 and FSE+, by the project CNS2024-154818 with funding by MICIU/AEI /10.13039/501100011033, by the project CIPROM/2022/66 with funding by the Generalitat Valenciana, and by the Ministry of Economic Affairs and Digital Transformation of the Spanish Government through the QUANTUM ENIA Project call—QUANTUM SPAIN Project, by the European Union through the Recovery, Transformation and Resilience Plan—NextGenerationEU within the framework of the Digital Spain 2026 Agenda, and by the CSIC Interdisciplinary Thematic Platform (PTI+) on Quantum Technologies (PTI-QTEP+). This work is supported through the project CEX2023-001292-S funded by MCIU/AEI.

\bibliographystyle{apsrev4-1} 
\bibliography{Reference}

\clearpage
\newpage

\begin{widetext}

\section{Supplemental material for ``Metrological Sensitivity beyond Gaussian Limits with Cubic Phase States''}

\section{I.\quad Quantum Fisher information}
\subsection{A. Analytical expression of QFI in rotation sensing}

A cubic phase state is generated by applying the cubic phase gate  $\hat{C}(r)=e^{i r \hat{x}^3}$, characterized by the cubicity parameter $r$, to a squeezed vacuum state that is squeezed along the $p$-axis:
\begin{align}\label{SMeq:cubicphase}
|\psi (r,s) \rangle &= \hat{C}(r) \hat{S}(\xi) |0\rangle \nonumber \\
&= e^{i r \hat{x}^3 } e^{ -s \left( \hat{a}^2 -\hat{a}^{\dagger 2} \right)/2 } |0\rangle.
\end{align}
Here, $\hat{S}(\xi)=\exp{\left[ \frac{1}{2} \left( \xi^* \hat{a}^2 -\xi \hat{a}^{\dagger 2} \right) \right]} $ is the squeezing operator, where $\xi=s e^{i\varphi}$, with squeezing strength $s$ and squeezing angle $\varphi$. In our case, we consider momentum squeezing, corresponding to $\varphi=\pi$, i.e., $\xi=-s$. 

The wave function of the states in Eq.~\eqref{SMeq:cubicphase} is
\begin{align}
\psi_x (r,s) &= \pi^{-1/4}  e^{-s/2} e^{-\frac{x^2}{2 e^{2s}}} e^{irx^3}.
\end{align}
Using $\hat{p}=-i\partial_x$, we first calculate the expectation values $\langle \hat{n} \rangle$ and $\langle \hat{n}^2 \rangle$ as
\begin{align}\label{SMeq:n}
\langle \hat{n} \rangle &= \int_{-\infty}^{\infty} dx \left(  \pi^{-1/4}  e^{-s/2} e^{-\frac{x^2}{2 e^{2s}}} e^{-irx^3} \right) \frac{1}{2} \left( x^2 -\partial_x^2 -1 \right) \left(  \pi^{-1/4}  e^{-s/2} e^{-\frac{x^2}{2 e^{2s}}} e^{irx^3} \right) \nonumber \\
&= \sinh^2{(s)} +\frac{27}{8} e^{4s} r^2, \\
\langle \hat{n}^2 \rangle &= \int_{-\infty}^{\infty} dx \left(  \pi^{-1/4}  e^{-s/2} e^{-\frac{x^2}{2 e^{2s}}} e^{-irx^3} \right) \frac{1}{4} \left( x^2 -\partial_x^2 -1 \right)  \left( x^2 -\partial_x^2 -1 \right) \left(  \pi^{-1/4}  e^{-s/2} e^{-\frac{x^2}{2 e^{2s}}} e^{irx^3} \right) \nonumber \\
&= \frac{ 1 }{ 64  } e^{ -4s} \left[ 12-16 e^{2s} +8e^{4s} +540 e^{10s} r^2 +8505 e^{12s} r^4 +e^{8s} \left(12-216 r^2  \right) +4e^{6s} \left( -4+45 r^2 \right) \right].
\end{align}
For pure states, the quantum Fisher information (QFI) is equivalent to four times of the variance of the phase generator, $F_Q[\psi (r,s),\hat{n}] = 4\text{Var}[\psi (r,s),\hat{n}]$. Thus, we obtain the analytical expressions of QFI in terms of $s$ and $r$ as
\begin{align}\label{SMeq:QFIscaling}
F_Q(r,s) &= \frac{1}{2} \left( -2+ e^{-4s} +e^{4s} +9 e^{2s} r^2 +54 e^{6s} r^2 +972 e^{8s} r^4 \right) .
\end{align}
If we eliminate $r$ using Eq.~\eqref{SMeq:n}, the QFI can be rewritten as a function of $n$ and $s$
\begin{align}\label{SMeq:QFIscaling}
F_Q(n,s) &= \frac{128}{3} n^2 + \left[ \left( \frac{4}{3}e^{-2s} +8e^{2s} \right) -\frac{256}{3} \sinh^2{(s)}  \right] n + \left[ \left( \cosh{(4s)} -1 \right) -\left( \frac{4}{3} e^{-2s} +8e^{2s} \right) \sinh^2{(s)} +\frac{128}{3}\sinh^4{(s)} \right],
\end{align}
where $n=\langle \hat{n}\rangle_{\psi (r,s)}$ denotes the population.

For $r=0$, the state in Eq.~\eqref{SMeq:cubicphase} returns to a squeezed vacuum state, which is known to provide the maximum QFI among Gaussian probe states in rotation sensing tasks~\cite{SahelNJP2025}. The expectation values $\langle \hat{n}\rangle$ and $\langle \hat{n}^2 \rangle$ for a squeezed vacuum states are
\begin{align}
\langle \hat{n} \rangle_{\psi (0,s)} &= \sinh^2{(s)} , \\
\langle \hat{n}^2 \rangle_{\psi (0,s)} &= \frac{1}{2} \sinh^2{(s)} \left( 1+3\cosh{(2s)} \right) , 
\end{align}
from which the corresponding QFI follows as
\begin{align}
F_Q[\psi (0,s),\hat{n}] &= \cosh{(4s)} - 1 \nonumber \\
&= 8 n \left( n +1 \right) .
\end{align}

\subsection{ B. Optimal parameters and their asymptotic values in large particle limit }

\subsubsection{1. Optimal parameters $s_{\text{opt}}$ and $r_{\text{opt}}$ }

After obtaining the analytical expression for the QFI of cubic phase state in Eq.~\eqref{SMeq:QFIscaling}, we now ask what the maximum sensitivity can be achieved for a given article number $n$, by optimizing the squeezing strength $s_{\text{opt}}$. The optimization problem can be solved as follows
\begin{align} \label{SMeq:OptimizationConds}
\frac{\partial F_Q(n,s)/n  }{ \partial s } &= 0, \\ \label{SMeq:OptimizationConds2}
\frac{\partial^2 F_Q(n,s)/n  }{ \partial s^2 } & <0. 
\end{align}
We first calculate
\begin{align}\label{SMeq:EqTarget}
\frac{\partial F_Q(n,s)/n  }{ \partial s } = \frac{1}{3n} e^{-4s} \left[ -34 +14e^{8s} +60 e^{2s} \left( 1+2n \right) -40 e^{6s} (1+2n) \right] & = 0.
\end{align}

Introducing the shorthand $A= 1+2 n$, we define $y= e^{2s} -\frac{5A}{7}$, $p = -\frac{150}{49} A^2$, $q = -\frac{1000}{343} A^3 +\frac{30}{7} A$ and $l = -\frac{1875}{2401} A^4 +\frac{150}{49} A^2 -\frac{17}{7}$, which allows us to reexpress Eq.~\eqref{SMeq:EqTarget} in a simpler form
\begin{align}
y^4 + py^2 +qy +l =0.
\end{align}
Using Ferrari's method, we rewrite this equation as
\begin{align}\label{SMeq:yfunction}
y^4 +py^2 +qy +l = \left( y^2 +m \right)^2 -(Cy+D)^2 =0.
\end{align}
By comparing the coefficients on the left- and right-hand sides, we obtain 
\begin{align}
p &= 2m -C^2 , \\
q &= -2CD , \\
l &= m^2 -D^2.
\end{align}
We can then rewrite the condition $m^2-D^2=l$ as
\begin{align}
2m^3 -pm^2 -2lm+ \left( pl -\frac{q^2}{4} \right) =0.
\end{align}
The solution $m^*$ satisfying this equation is given by
\begin{align}
m^* &= m^*_{1} (n) + m^*_{2} (n) + m^*_{3} (n), \\
m^*_{1} (n) &= -\frac{25}{49} \left( 1+4n+4n^2 \right), \\
m^*_{2} (n) &= -\frac{ 7\left(-\dfrac{372}{49} - \dfrac{7200n}{49} - \dfrac{7200n^{2}}{49}\right) }{ 6^{4/3}  } S(n)^{-1/3}, \\
m^*_{3} (n) &= \frac{1}{7\cdot 6^{2/3}} S(n)^{1/3},
\end{align}
where 
\begin{align} \label{SMeq:Sn}
S(n) &= -1125 -4500n -4500 n^2 +\sqrt{3} \sqrt{Q(n)} , \\ \label{SMeq:Qn}
Q(n) &= -54781 - 24301800n - 553231800n^{2}
- 4513860000n^{3} - 10896930000n^{4}
- 10368000000n^{5} - 3456000000n^{6}.
\end{align}

Substituting $m^*$ in Eq.~\eqref{SMeq:yfunction}, we obtain 
\begin{align}\label{SMeq:ystar}
y^* 
&= \frac{1}{2} \left[ \sqrt{2m^* -p} -\left( 2m^* -p  \right) -4(m^* + \frac{q}{2\sqrt{2m^*-p}} ) \right],
\end{align}
from which the optimal squeezing strength follows as
\begin{align}
s_{\text{opt}} &= \frac{1}{2} \log{\left[ y^* +\frac{5(1+2n)}{7} \right]}.
\end{align}
Here, $y^*$ is chosen as one of the four roots of Eq.~\eqref{SMeq:yfunction}, such that the corresponding squeezing strength follows $s_{\text{opt}}>0$ and both conditions in Eq.~\eqref{SMeq:OptimizationConds2} are satisfied.

\subsubsection{2. Asymptotic values in the limit of large $n$ }

In the limit of large $n$, we first rewrite $Q(n)$ in Eq.~\eqref{SMeq:Qn} as
\begin{align}
Q(n) &= - 3456000000 n^6 \left[ 1+ \frac{3}{n} +\mathcal{O}(n^{-2}) \right]  .
\end{align}
Defining $\mathcal{D}\equiv \sqrt{3} \sqrt{3456000000}$, we obtain $ \sqrt{3} \sqrt{Q(n)} = i \mathcal{D} n^3 \left[ 1+ \frac{3}{2} n^{-1} +\mathcal{O}(n^{-2}) \right]  $.
Accordingly, $S(n)$ in Eq.~\eqref{SMeq:Sn} can be expressed as
\begin{align}
S(n) 
&= i \mathcal{D} n^3 \left[ 1 + \left( \frac{3}{2} - \frac{4500}{i \mathcal{D}} \right) \frac{1}{n} +\mathcal{O}(n^{-2}) \right].
\end{align}
Introducing $\mathcal{K} = (i\mathcal{D})^{1/3} =\mathcal{D}^{1/3} e^{i\pi/6}$, we obtain
\begin{align}
S(n)^{1/3} &= \mathcal{K} n \left[ 1 +\frac{1}{3} \left( \frac{3}{2} - \frac{4500}{i \mathcal{D}} \right) n^{-1} +\mathcal{O}(n^{-2})  \right], \\
S(n)^{-1/3} &= \mathcal{K}^{-1} n^{-1} \left[ 1 -\frac{1}{3} \left( \frac{3}{2} - \frac{4500}{i \mathcal{D}} \right) n^{-1} +\mathcal{O}(n^{-2})  \right]
\end{align}
From $S(n)^{1/3}$ and $S(n)^{-1/3}$, $m^*_2$ and $m^*_3$ for $n\rightarrow \infty$ can be approximated as
\begin{align}
m^*_2 
& \approx  \frac{ 7200 }{7\cdot 6^{4/3} \mathcal{K}} n, \\
m^*_3 
& \approx \frac{1}{ 7\cdot 6^{2/3}} \mathcal{K} n .
\end{align}
The asymptotic value $m^*_{n\rightarrow\infty}$ is then given by
\begin{align}
m^*_{n\rightarrow\infty} &= m^*_1 +  m^*_2  + m^*_3 \nonumber \\
&\approx -\frac{25}{ 49 } \left( 1 + 4n +4n^2 \right) + \frac{ 7200 }{7\cdot 6^{4/3} \mathcal{K}} n + \frac{1}{ 7\cdot  6^{2/3}} \mathcal{K} n \nonumber \\
&= - \frac{100}{ 49} n^2 + \frac{10}{49}\left( -10+7\sqrt{6} \right) n,
\end{align}
which yields $y_{n\rightarrow \infty}^*$ in Eq.~\eqref{SMeq:ystar} as
\begin{align}
y_{n\rightarrow \infty}^* 
& \approx -\frac{10}{7}n - \frac{5}{7} +\sqrt{ \frac{3}{2} } .
\end{align}
Finally, we obtain the asymptotic value of the optimal squeezing strength,
\begin{align}
s^{n\rightarrow \infty}_{\text{opt}} &= \frac{1}{2} \log{\left[ y_{n\rightarrow \infty}^* +\frac{5(1+2n)}{7} \right]} \nonumber \\
& \approx \frac{1}{2} \log{\left( \frac{\sqrt{6}}{2}  \right)} \nonumber \\
& \approx 0.101366,
\end{align}
which corresponds to $0.880454$ dB of squeezing.

The asymptotic value of the optimal optimal cubic strength is then obtained from Eq.~\eqref{SMeq:n}, which reads
\begin{align}
|r^{n\rightarrow \infty}_{\text{opt}}| &= \frac{ n-\sinh^2{ \left( s^{n\rightarrow \infty}_{\text{opt}} \right)} }{ \frac{27}{8} e^{4s} } , \\
&\approx \frac{4 \sqrt{n}}{9}.
\end{align}

\clearpage
\subsection{C. Quantum Fisher information in displacement sensing}

We also investigate the metrological performance of cubic phase states in displacement sensing. A parameter $d$ is imprinted on the probe state $|\psi(r,s)\rangle$ via $e^{-i\hat{G}d}|\psi(r,s)\rangle$, where the phase generator is $\hat{G}=\cos{(\phi)} \hat{x} + \sin{(\phi)} \hat{p}$, resulting a displacement in the phase space along a direction associated with $\phi$. The optimal QFI is obtained by optimizing over the displacement direction, $F_Q[|\psi(r,s)\rangle,\hat{G}] = \max_{\phi} 4 \text{Var}[|\psi(r,s)\rangle,\hat{G}(\phi)]$. In Fig.~\ref{FigSM_Displacement}, we plot $F_Q/n$ as a function of squeezing strength $s$. It is found that, for cubicity $r>0$, the sensitivity does not surpass that of squeezed vacuum states ($r=0$, black line), indicating the cubic phase state can not outperform Gaussian states in displacement sensing. 
\begin{figure}[h]
    \centering
    \includegraphics[width=90mm]{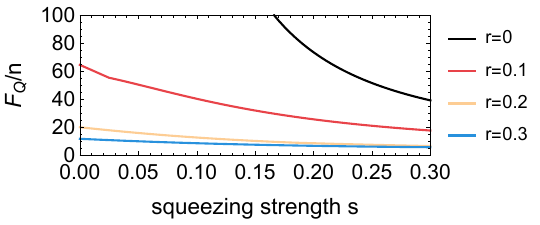}
    \caption{\textbf{The sensitivity of cubic phase states in displacement sensing.} $F_Q/n$ as a function of squeezing strength $s$ for different cubicity $r$. At $r=0$, it becomes squeezed vacuum states. }
    \label{FigSM_Displacement}
\end{figure}

Ref.~\cite{MatteoRPP2025} discussed displacement sensing under different assumptions. If the phase $\phi$ of the displacement is known and aligned with the most sensitive quadrature, the ultimate sensitivity can be saturated by  optimally oriented squeezed vacuum states. By contrast, if the displacement direction is unknown, the sensitivity is evaluated along the least sensitive quadrature. This optimal worst-case sensitivity is achieved by non-Gaussian probes such as Fock states and compass states, and no Gaussian states can attain this ultimate bound. Therefore, although squeezed vacuum states are the optimal probes to achieve the maximum sensitivity in displacement sensing, non-Gaussian states show advantage in certain scenarios. Such non-Gaussian states have already been prepared in CV experiments~\cite{FabianNC2019,RahmanPRL2025} and exploited in metrological tasks~\cite{FabianNC2019}.

\section{II.\quad Nonlinear squeezing parameters}
\subsection{A. The method of moments }
In rotation sensing, a phase shift $\theta$ is encoded to a probe states $\rho$ via the unitary interaction $e^{-i\hat{n}\theta} \rho e^{i\hat{n}\theta} $. Using the method of moment, the uncertainty in estimating the phase shift from the outcomes of a measurement $\hat{M}$ is given by
\begin{align}\label{SMeq:uncertainty}
\text{Var}[\theta_{\text{est}}] &= \frac{\text{Var}[\rho,\hat{M}]}{ |\langle[\hat{n},\hat{M}]\rangle_{\rho}|^2 }.
\end{align}
We now consider a set of observables $\mathbf{M}=\left( \hat{M}_1, \cdots, \hat{M}_l \right)$, so that any accessible measurement can be written as a linear combination $\hat{M}=\vec{m} \cdot \mathbf{M}$. The inverse of the squeezing parameter in Eq.~\eqref{SMeq:uncertainty} can then be expressed as
\begin{align}
\chi^{-2}[\rho, \hat{n}, \hat{M}] = \frac{ |  \mathbf{C}[\rho,\hat{n},\mathbf{M}] \vec{m} |^2}{ \vec{m}^T \boldsymbol{\Gamma}[\rho,\mathbf{M}] \vec{m} },
\end{align}
where $\mathbf{C}[\rho,\hat{n},\mathbf{M}]$ is the commutator vector with elements $\left( \mathbf{C}[\rho,\hat{n},\mathbf{M}] \right)_{j} = -i \langle [ \hat{n}, \hat{M}_j] \rangle_\rho$, and $\boldsymbol{\Gamma}[\rho,\mathbf{M}]$ is the covariance matrix with elements $\left( \boldsymbol{\Gamma}[\rho,\mathbf{M}] \right)_{jk} = \text{Cov}(\hat{M}_j,\hat{M}_k)_{\rho}$. 

If a moment matrix is introduced,
\begin{align}
\mathcal{M}[\rho,\hat{n},\mathbf{M}] = \mathbf{C}[\rho,\hat{n},\mathbf{M}] \boldsymbol{\Gamma}[\rho,\mathbf{M}]^{-1}  \mathbf{C}[\rho,\hat{n},\mathbf{M}]^T,
\end{align}
the maximum achievable sensitivity over all linear combination $\vec{m}$ can be obtained following~\cite{ManuelPRL2019}
\begin{align}
\chi^{-2}_{\text{opt}}[\rho,\hat{n},\mathbf{M}] &= \max_{\vec{m}} \chi^{-2}[\rho, \hat{n}, \hat{M}] \nonumber \\
&= \mathcal{M}[\rho,\hat{n},\mathbf{M}],
\end{align}
with the corresponding optimal vector
\begin{align}
\vec{m}_{\text{opt}} &= \alpha \boldsymbol{\Gamma}[\rho,\mathbf{M}]^{-1} \mathbf{C}[\rho,\hat{n},\mathbf{M}]^T,
\end{align}
where $\alpha$ is a normalization constant. For better comparing sensitivity for different population $n$, we define the squeezing coefficient normalized by $n$ as
\begin{align}\label{SMeq:xi2}
\xi^{-2} [\rho,\hat{n},\mathbf{M}] &= \frac{\chi^{-2}_{\text{opt}} [\rho,\hat{n},\mathbf{M}] }{n}.
\end{align}

\subsection{B. Analytical expression for nonlinear squeezing parameters }

The sensitivity in Eq.~\eqref{SMeq:xi2} depends on the choice of the measurement $\hat{M}$. We now introduce a family $\mathbf{M}_{(k)}$ containing quadrature operators up to $k-$order. For $k=1$, the set contains linear measurements, $\mathbf{M}_{(1)} = \left( \hat{x},\hat{p} \right)$, yielding the linear squeezing coefficient $\xi^{-2}_{(1)}=\xi^{-2}_{(1)} [\rho,\hat{n},\mathbf{M}_{(1)}]$. For $k>1$, higher order observables are involved in $\mathbf{M}_{(k)}$, leading to nonlinear squeezing coefficients $\xi^{-2}_{(k)}$.

In the case of pure cubic phase states, some elements of the commutation vector are zero, so that the effective $k-$th order measurement sets are reduced to the following sets that lead to nonzero contributions:
\begin{align}\label{SMeq:Mset1}
\mathbf{M}_{(1)} &= \{\hat{x}\} = \{ \hat{\mathcal{X}}_{1} \} , \\ \label{SMeq:Mset2}
\mathbf{M}_{(2)} &= \{ \mathbf{M}_{(1)}, \frac{1}{2}\left( \hat{x}\hat{p}+\hat{p}\hat{x} \right) \} = \{ \hat{\mathcal{X}}_{1}, \hat{\mathcal{X}}_{2} \} , \\ \label{SMeq:Mset3}
\mathbf{M}_{(3)} &= \{ \mathbf{M}_{(2)}, \hat{x}^3, \frac{1}{3}\left( \hat{x}\hat{p}\hat{p}+\hat{p}\hat{x}\hat{p} + \hat{p}\hat{p}\hat{x} \right) \} = \{ \hat{\mathcal{X}}_{1}, \hat{\mathcal{X}}_{2}, \hat{\mathcal{X}}_{3},\hat{\mathcal{X}}_{4}\} , \\ \label{SMeq:Mset4}
\mathbf{M}_{(4)} &= \{ \mathbf{M}_{(3)},  \frac{1}{4} \left( \hat{x}^3 \hat{p} + \hat{x}^2 \hat{p}\hat{x} + \hat{x} \hat{p}\hat{x}^2 +\hat{p}\hat{x}^3 \right), \frac{1}{4} \left( \hat{x} \hat{p}^3 + \hat{p}\hat{x} \hat{p}^2 + \hat{p}^2 \hat{x}\hat{p} +\hat{p}^3\hat{x} \right) \} = \{ \hat{\mathcal{X}}_{1}, \hat{\mathcal{X}}_{2}, \hat{\mathcal{X}}_{3},\hat{\mathcal{X}}_{4} , \hat{\mathcal{X}}_{5},\hat{\mathcal{X}}_{6}\}.
\end{align}
Here, we use the shorthand $\hat{\mathcal{X}}_i$ to denote the effective operators.

To construct the covariance matrix $\boldsymbol{\Gamma}[\rho,\mathbf{M}_{(k)}]$, we derive analytical expressions for its elements $\left( \boldsymbol{\Gamma}[\rho,\mathbf{M}_{(k)}]\right)_{jk}=\text{Cov}\left( \hat{\mathcal{X}}_j, \hat{\mathcal{X}}_k \right)$,
\begin{align}
\text{Cov}\left( \hat{\mathcal{X}}_1,\hat{\mathcal{X}}_1 \right) &= \frac{1}{2} e^{2s} , \\
\text{Cov}\left( \hat{\mathcal{X}}_1,\hat{\mathcal{X}}_2 \right) &=  \frac{9}{4} r e^{4s} , \\
\text{Cov}\left( \hat{\mathcal{X}}_1,\hat{\mathcal{X}}_3 \right) &=  \frac{3}{4} e^{4s} , \\
\text{Cov}\left( \hat{\mathcal{X}}_1,\hat{\mathcal{X}}_4 \right) &=  \frac{1}{8} \left( 135 r^2 e^{6s} +2 \right) , \\
\text{Cov}\left( \hat{\mathcal{X}}_1,\hat{\mathcal{X}}_5 \right) &=  \frac{45}{8} r e^{6s} , \\
\text{Cov}\left( \hat{\mathcal{X}}_1,\hat{\mathcal{X}}_6 \right) &=  \frac{21}{16} r e^{2s} \left( 135 r^2 e^{6s} +2 \right) , \\
\text{Cov}\left( \hat{\mathcal{X}}_2,\hat{\mathcal{X}}_2 \right) &= \frac{1}{8} \left( 135 r^2 e^{6s} +4 \right) ,\\
\text{Cov}\left( \hat{\mathcal{X}}_2,\hat{\mathcal{X}}_3 \right) &=  \frac{45}{8} r e^{6s} , \\
\text{Cov}\left( \hat{\mathcal{X}}_2,\hat{\mathcal{X}}_4 \right) &=  \frac{27}{16} r e^{2s} \left( 105 r^2 e^{6s} +2 \right), \\
\text{Cov}\left( \hat{\mathcal{X}}_2,\hat{\mathcal{X}}_5 \right) &=  \frac{3}{16} e^{2s}\left( 315 r^2 e^{6s} +4 \right) , \\
\text{Cov}\left( \hat{\mathcal{X}}_2,\hat{\mathcal{X}}_6 \right) &=  \frac{3}{32} e^{-2s} \left(  45 r^2 e^{6s} \left( 567 r^2 e^{6s} +10\right) +8 \right) , \\
\end{align}
\begin{align}
\text{Cov}\left( \hat{\mathcal{X}}_3,\hat{\mathcal{X}}_3 \right) &= \frac{15}{8} e^{6s} , \\
\text{Cov}\left( \hat{\mathcal{X}}_3,\hat{\mathcal{X}}_4 \right) &=  \frac{3}{16} e^{2s} \left( 315 r^2 e^{6s} -2 \right), \\
\text{Cov}\left( \hat{\mathcal{X}}_3,\hat{\mathcal{X}}_5 \right) &=  \frac{315}{16} r e^{8s} , \\
\text{Cov}\left( \hat{\mathcal{X}}_3,\hat{\mathcal{X}}_6 \right) &=  \frac{45}{32} r e^{4s} \left( 567 r^2 e^{6s}-2 \right) , \\
\text{Cov}\left( \hat{\mathcal{X}}_4,\hat{\mathcal{X}}_4 \right) &=  \frac{1}{32} e^{-2s} \left[ 27 r^2 e^{6s} \left( 2835 r^2 e^{6s} +52 \right) +28 \right], \\
\text{Cov}\left( \hat{\mathcal{X}}_4,\hat{\mathcal{X}}_5 \right) &=  \frac{9}{32} r e^{4s} \left( 2835 r^2 e^{6s} +26 \right) , \\
\text{Cov}\left( \hat{\mathcal{X}}_4,\hat{\mathcal{X}}_6 \right) &=  \frac{3}{64} r \left[ 945 r^2 e^{6s} \left( 891 r^2 e^{6s} +16 \right) +124 \right] , \\
\text{Cov}\left( \hat{\mathcal{X}}_5,\hat{\mathcal{X}}_5 \right) &= \frac{21}{32} e^{4s} \left( 405 r^2 e^{6s} +4 \right), \\
\text{Cov}\left( \hat{\mathcal{X}}_5,\hat{\mathcal{X}}_6 \right) &=  \frac{3}{64} \left[ 45 r^2 e^{6s} \left( 6237 r^2 e^{6s} +50 \right) -8 \right] , \\
\text{Cov}\left( \hat{\mathcal{X}}_6,\hat{\mathcal{X}}_6 \right) &= \frac{3}{128} e^{-4s} \left[ 15 r^2 e^{6s} \left( 2457 r^2 e^{6s} (891 r^2 e^{6s}+16) +356 \right) +112 \right].
\end{align}
The remaining nonzero elements are determined by the symmetry property of the covariance matrix, $\boldsymbol{\Gamma}[\rho,\mathbf{M}] = \boldsymbol{\Gamma}[\rho,\mathbf{M}]^T$.

Similarly, to construct the commutator vector $\mathbf{C}[\rho,\hat{n},\mathbf{M}_{(k)}]$, we obtain analytical expressions for its elements $\left( \mathbf{C}[\rho,\hat{n},\mathbf{M}_{(k)}]\right)_{j}= -i \langle [\hat{n},\hat{\mathcal{X}}_j]\rangle_{\rho}$,
\begin{align}
-i \langle [ \hat{n}, \hat{\mathcal{X}}_1 ] \rangle &= -\frac{3}{2} r e^{2s}, \\
-i \langle [ \hat{n}, \hat{\mathcal{X}}_2 ] \rangle &= -\frac{27}{4} r^2 e^{4s} +\sinh{(2s)} ,\\
-i \langle [ \hat{n}, \hat{\mathcal{X}}_3 ] \rangle &= -\frac{27}{4} r e^{4s}, \\
-i \langle [ \hat{n}, \hat{\mathcal{X}}_4 ] \rangle &= -\frac{3}{8} r \left( 135 r^2 e^{6s} -12 e^{4s} +2 \right) ,\\
-i \langle [ \hat{n}, \hat{\mathcal{X}}_5 ] \rangle &= -\frac{3}{8}  \left( 135 r^2 e^{6s} -2 e^{4s} +2 \right) ,\\
-i \langle [ \hat{n}, \hat{\mathcal{X}}_6 ] \rangle &= -\frac{3}{16} e^{-4s} \left[ 15 r^2 e^{6s} \left( 189 r^2 e^{6s} -18 e^{4s} +4 \right) - 4e^{4s} +4 \right].
\end{align}

Using the expressions for the covariance matrix and the commutator vector, we construct the $k$-th moment matrix $\mathcal{M}_{(k)}$. The corresponding optimal squeezing parameters $\chi^{-2}_{(k)}=\mathcal{M}_{(k)}$ can then be expressed in terms of $r$ and $s$,  
\begin{align}
\chi_{(1)}^{-2} &= \frac{9}{2} r^2 e^{2s}   ,\\
\chi_{(2)}^{-2} &= \frac{9}{2} r^2 e^{2s} + \frac{ 4\sinh^2{(2s)}}{ 27 r^2 e^{6s} +2 }  ,\\
\chi_{(3)}^{-2} &= \frac{1}{10} \left[ 9 r^2 e^{2s} \left( 29 -\frac{192}{ 45 r^2 e^{6s} +8} \right) +3645 r^4 e^{8s} +270 r^2 e^{6s} + 5 e^{4s} +5e^{-4s} -10 \right]  ,\\
\chi_{(4)}^{-2} &= \frac{1}{2} \left( -2+ e^{-4s} +e^{4s} +9 e^{2s} r^2 +54 e^{6s} r^2 +972 e^{8s} r^4 \right) ,
\end{align}
which can equivalently be written as function of $n$ and $s$
\begin{align}
\chi_{(1)}^{-2} 
&= \frac{4}{3} e^{-2s} \left( n -\sinh^2{(s)} \right)  ,\\
\chi_{(2)}^{-2} 
&= \frac{4}{3} e^{-2s} \left( n -\sinh^2{(s)} \right) +\frac{4\sinh^2{(2s)} }{ 8 e^{2s} \left( n -\sinh^2{(s)} \right) +2 } ,\\
\chi_{(3)}^{-2} 
&=  32 \left( n -\sinh^2{(s)} \right)^2 + 8 e^{2s} \left( n -\sinh^2{(s)}  \right) +\frac{4}{15} e^{-2s} \left( n -\sinh^2{(s)} \right) \left( 29-\frac{ 72 }{ 5e^{2s} \left( n -\sinh^2{(s)} \right) +3 }  \right) +\cosh{(4s)} -1 ,\\
\chi_{(4)}^{-2} 
&= \frac{128}{3} n^2 + \left[ \left( \frac{4}{3}e^{-2s} +8e^{2s} \right) -\frac{256}{3} \sinh^2{(s)}  \right] n + \left[ \left( \cosh{(4s)} -1 \right) -\left( \frac{4}{3} e^{-2s} +8e^{2s} \right) \sinh^2{(s)} +\frac{128}{3}\sinh^4{(s)} \right].
\end{align}
We find that the sensitivity revealed by the third- and the fourth-order parameters are significantly larger than those of first and second order, as they show a factor of $n$ improvement in the leading term. In the limit of large $n$, the third-order parameter scales as $\chi_{(3)}^{-2} \sim 32 n^2 $, while the fourth-order parameter scales as $\chi_{(4)}^{-2} \sim \frac{128}{3}n^2 $. Furthermore, we observe that the fourth-order squeezing parameter coincides with the QFI in Eq.~\eqref{SMeq:QFIscaling} for any $r$ and $s$, indicating that the optimal measurement observables saturating the quantum Cram$\acute{\text{e}}$r-Rao bound are contained in $\mathbf{M}_{(4)}$.

\subsection{C. Proof of locally unbiased estimator}
In our scenarios, we aim to estimate a small phase shift around zero, $\theta \simeq 0$. Given a measurement observable $\hat{M}$, we denote the outcome from a single shot by $a_i$, and define the estimator 
\begin{align}
\theta_{\text{est}}(a_i) &= \frac{a_i}{ \frac{\partial \langle \hat{M} (\theta) \rangle }{ \partial \theta } \Big|_{\theta=0} }.
\end{align}
For $m$ independent measurements, the mean value of the outcomes is
\begin{align}
\overline{a}_m &= \frac{1}{m} \sum_{i=1}^m a_i.
\end{align}
Using $\mathbbm{E}[\overline{a}_m] = \langle \hat{M} (\theta) \rangle$, we thus obtain 
\begin{align}
\mathbbm{E}[\theta_{\text{est}}  ]_{\theta=0} &= \frac{ \langle \hat{M} (\theta=0) \rangle }{ \frac{\partial \langle \hat{M} (\theta) \rangle }{ \partial \theta } \Big|_{\theta=0} }.
\end{align}

The $k$th-order measurements $\hat{M}_k$ to yield the nonlinear squeezing coefficients $\xi_{(k)}^{-2}$ are 
\begin{align}
\hat{M}_k &= \vec{m}^{(k)}\cdot \mathbf{M}_{(k)} = \sum_{i} m_i^{(k)} \hat{\mathcal{X}}_i.
\end{align}
in SM~\cite{SM} Sec. II B, we have obtained analytical expressions 
\begin{align}
\langle \hat{\mathcal{X}}_i \rangle &= \langle \hat{\mathcal{X}}_i(\theta=0) \rangle = 0,
\end{align}
for all observables $i=\{1,\cdots,6\}$. Hence any linear combination of them satisfies
\begin{align}
    \langle \hat{M}_k (\theta=0) \rangle &= 0.
\end{align}
Using $\frac{\partial \langle \hat{M}_k (\theta) \rangle }{ \partial \theta } \Big|_{\theta=0}  \neq 0$, we obtain 
\begin{align}
\mathbbm{E}[\theta_{\text{est}}]_{\theta=0} &= \frac{ \langle \hat{M}_k (\theta=0) \rangle }{ \frac{\partial \langle \hat{M}_k (\theta) \rangle }{ \partial \theta } \Big|_{\theta=0} } = 0, \\
\frac{ \partial \mathbbm{E}[\theta_{\text{est}}] }{ \partial \theta } \Big|_{\theta=0} &=  1.
\end{align}
In conclusion, all the estimators we used in nonlinear squeezing coefficients are locally unbiased at $\theta=0$.

\subsection{D. Comparison with ``sub-Heisenberg'' states}

In CV phase sensing, it is possible to design pure states with arbitrarily large number fluctuations at fixed population $n$, thereby achieving unbounded sensitivity. Since these states may exceed the quadratic scaling of the $F_Q$ that is sometimes referred to as ``Heisenberg scaling'', they have raised considerable interest as potential strategies to achieve ``sub-Heisenberg'' estimation error. 

One typical example is the class of states introduced by Shapiro, Shepard, and Wong (SSW) in Ref.~\cite{ShapiroPRL1989}, 
\begin{align}\label{SMeq:SSWstate}
|\psi_{\text{SSW}} \rangle &= A \sum_{k=0}^M \frac{e^{ik\phi}}{k+1} |k\rangle,
\end{align}
where $M$ is a positive integer to truncate the series and $A$ is a normalization constant. For a large finite population $n$, these SSW states exhibit an exponential explosion in the number fluctuation $\Delta n \sim \text{exp}\left( \pi^2 n/12 \right)$, implying an unbounded QFI. The sensitivity obtained from the reciprocal peak likelihood estimator shows the ``sub-Heisenberg'' scaling, $\delta \Phi   = \frac{2\pi A^2}{ (n+1)^2 } $. However, this result is reassessed in Refs.~\cite{BraunsteinPRL1992,LanePRA1993}, where it shows that such extreme sensitivity arises only at the cost of an enormous number of detection events. To be concrete, the phase distribution
\begin{align}
P_{\text{SSW}}(\Phi|\phi) &= \frac{A^2}{ 2\pi } \Big| \sum_{k=0}^M \frac{e^{-ik(\Phi-\phi)}}{ k+1 } \Big|^2
\end{align}
possesses two unusual features: a sharpened central peak at $\Phi=\phi$ together with broad, slowing decaying tails. For $n \gtrsim 2$, the distribution is a heavy-tailed envelop away from the central region, $P_{\text{SSW}}(\Phi|\phi) \simeq \frac{A}{2\pi} \abs{ \text{ln} \left( 1- e^{-i(\Phi-\phi)} \right) }^2 $. While in the central region, the cutoff $M$ softens the logarithmic singularity into a Gaussian peak with height $(n+1)^2/(2\pi A)$ and a narrow width $\sim 1/M$. Thus, almost all of the probability weight lies in the broad logarithmic tails. Refs.~\cite{BraunsteinPRL1992,LanePRA1993} considered the maximum-likelihood estimation (MLE) and performed Monte Carlo simulations, they found that for moderate sample size the estimator only probes heavy-tailed part of the distribution, therefore is insensitive. Only when the number of measurements are large enough, the statistics can be found in the tiny peak region, so that the estimator can yield the sensitivity approaching the Cram$\acute{\text{e}}$r-Rao bound. Consequently, After taking the cost of measurements into account, the SSW states do not exhibit any sub-Heisenberg advantage.

Another example is the superposition of vacuum and squeezed states discussed by Rivas and Luis (RL) in Ref.~\cite{RivasNJP2012}, 
\begin{align}
|\psi_{\text{RL}}\rangle &= \mu |0\rangle + \nu |\alpha, \xi \rangle,
\end{align}
where $|0\rangle$ and $|\alpha,\xi \rangle =\hat{D}(\alpha)\hat{S}(\xi)|0\rangle$ represent a vacuum state and a quadrature squeezed state, respectively. If one sets the squeezing parameter $\xi=s$ and the displacement amplitude $\alpha=\sinh(s)e^{i\pi/2}$, the QFI of the RL state becomes $F_Q=6n^2/|\nu|^2$. 
By fixing the particle $n$ and choosing an infinitely small $|\nu|\rightarrow 0$, one can make QFI arbitrarily large, potentially leading to ``sub-Heisenberg'' sensitivity. In the RL scheme, the phase $\phi$ is estimated from the quadrature observable $\hat{x}$. The probability distribution is
\begin{align}
P_{\text{RL}}(x|\phi)= |\mu \langle x|0\rangle + \nu \langle x|\alpha,\xi,\phi \rangle |^2,
\end{align}
where $|\alpha,\xi,\phi\rangle = e^{-i\phi \hat{n} }|\alpha,\xi \rangle$, and the vacuum is insensitive to rotation signal $e^{-i\phi \hat{n} }|0\rangle = |0\rangle$. At $\phi=0$,  Fisher information is given by $F(0) \approx 4n^2/\nu^2$, almost saturating the QFI. In the probability distribution of the measurement $\hat{x}$, the squeezed state contributes an extremely narrow central Gaussian peak, leading to the increased sensitivity depending on the weight factor $\nu$ and squeezing strength $s$, whereas the dominant vacuum produces a much broader Gaussian background that is insensitive to the phase. With smaller $\nu$, the peak becomes higher and narrower, but its integrated weight becomes vanishingly small: The probability for the statistics to contribute to the narrow peak region is $p\approx \nu/\sqrt{n}$. Consequently, measurement outcomes generally fall with high probability in the broad but insensitive background, whereas only few events will cover the highly sensitive peak. Detailed numerical analysis of the maximum likelihood estimator~\cite{LacaPRA2013} revealed that the ``sub-Heisenberg'' sensitivity is related to a strong bias of the estimator, rather than a genuine metrological advantage. After the bias correction, no violation of Heisenberg limit is found anymore. This bias is caused by the unusual feature of the statistics distribution: most of the statistics is located in the broad background provided by the vacuum states, which is invariant under rotation perturbation, leading to a phase-insensitive MLE distribution.

Both SSW states and RL states exhibit arbitrarily large number fluctuation at fixed population, which at the level of QFI suggests a ``sub-Heisenberg'' sensitivity. However, in both cases the corresponding measurement statistics have unusual features, such as a narrow, highly sensitive peak on top of a broad but insensitive background. Only the extremely rare measurement outcomes that fall within this narrow peak carry significant phase information, while the vast majority of detection events contribute negligibly and are effectively wasted. As a result, achieving the large QFI requires an excessively large number of measurement repetitions. Once the measurement number is taken account and estimator bias is corrected, no violation of Heisenberg limit is actually found. Recently, it was also shown that under realistic noise considerations, the Fisher information of RL-type states only exceeds Gaussian limits in a certain energy range~\cite{KiminarXiv2025}.

Our scheme based on cubic phase states is conceptually distinct from these ``sub-Heisenberg'' protocols. In our case, the number fluctuation still follows the quadratic, so-called Heisenberg scaling $\text{Var}(n)\sim n^2$, and so does the corresponding QFI, $F_Q\sim 42.7n^2$. We obtain a prefactor enhancement beyond all Gaussian probes, while maintaining the Heisenberg scaling, rather than introducing arbitrary large sensitivity scaling at fixed population. This makes our protocol robust to statistical analysis in practical implementations. 

The measurement statistics of cubic phase states neither exhibits a broad background nor narrow peaks. The wave function of a cubic phase state in $x$ representation is given by $\psi_x (r,s) = e^{irx^3} \psi_{\text{sq}}$, where $\psi_{\text{sq}}$ is the wave function of a squeezed vacuum state. Hence the corresponding modulus is $\abs{\psi_x(r,s)}^2=\abs{\psi_{\text{sq}}}^2$, which follows a Gaussian distribution with rapidly decaying tails. The cubic phase operator $e^{ir\hat{x}^3}$ only contributes a phase, which does not affect the fast decaying trend. The observables $\hat{M}\in \mathbf{M}_{(k)}$ are Hermitian polynomials of at most $k$-th order of quadrature measurements $\hat{x}$ and $\hat{p}$. Their outcome distribution can thus be viewed as a rapidly decaying $(x,p)$ distributions under a polynomial map $f(x,p)$. Therefore, the measurement statistics remain regular non-Gaussian distributions with fast decaying tails. This ensures that the majority of our detection events carry phase information. We also proved in SM~\cite{SM} Sec. II C that our estimator based on the method of moment is locally unbiased.

\section{III.\quad Robustness to noise }
In this section, we take realistic imperfections into account and analyze the impact of two typical noise sources on the metrological sensitivity of cubic phase states: population loss and detection noise.

\subsection{ A. Population loss }

We first investigate the effect of loss, which is a dominant source leading to decoherence in continuous variable systems. In the ideal case, a cubic phase state is a pure state generated by two sequential unitary operators, $|\psi\rangle = \hat{C}(r) \hat{S}(s)|0\rangle = e^{ i r \hat{x}^3 } e^{s\left( \hat{a}^{\dagger 2} - \hat{a}^2 \right)/2 } |0\rangle$. We can equivalently write the state in the form of the Hamiltonian evolutions, 
\begin{align}
|\psi \rangle &= e^{-i \hat{H}_2 t_2} e^{-i \hat{H}_1 t_1} |0\rangle
\end{align}
with 
\begin{align}
\hat{H}_1 &= i s' \left( \hat{a}^{\dagger 2} -\hat{a}^2 \right)/2 , \\
\hat{H}_2 &= -r' \hat{x}^3, 
\end{align}
Here we take $s'=s/t_1$ and $r'=r/t_2$, where $t_1, t_2$ denote the respective evolution times. In our calculations, we take $t_1=t_2=1$ as parameterized values, so that $s'$ and $r'$ coincide with the squeezing and cubicity parameters. The loss can be described by a jump operator $\hat{L}_k=\sqrt{\gamma_k} \hat{a}$ ($k=1,2$) in a Lindblad master equation,
\begin{align}
\frac{d \rho}{ d t_k} &= -i [\hat{H}_k, \rho] + \hat{L}_k \rho \hat{L}^\dagger_k - \frac{1}{2} \{ \hat{L}^\dagger_k \hat{L}_k, \rho \},
\end{align}
where we take equal loss rates, $\gamma_k=\gamma$. The state is initially prepared in a vacuum state $|0\rangle$, and then evolves sequentially according to the master equations with intrinsic Hamiltonians $\hat{H}_1$ and $\hat{H}_2$. Fig.~\ref{Fig3}(b) in the main text shows nonlinear squeezing coefficients as a function of the loss rate $\gamma$. We observe that the third- and fourth-order squeezing coefficients $\xi_{(3)}^{-2}$ and $\xi_{(4)}^{-2}$ decreases with the loss $\gamma t$, while the linear and second-order squeezing coefficients $\xi_{(1)}^{-2}$ and $\xi_{(2)}^{-2}$ exhibit a slight increase. This is because the loss reduces both unnormalized sensitivities $\chi_{(k)}^{-2}$ and the population $n$, the competition between the two parameters determines the change of the resulting normalized sensitivities $\xi_{(k)}^{-2}=\chi_{(k)}^{-2}/n$. As an intrinsic noise source, loss acts directly at the state level, changing the pure state to a mixed states and thereby reducing its QFI. For a mixed state, QFI is expressed as
\begin{align}
F_Q[\rho,\hat{n}]=\sum_{\substack{k,l\\ \text{s.t. } \lambda_k+\lambda_l>0 }}\frac{(\lambda_k-\lambda_l)^2}{(\lambda_k+\lambda_l)} |\bra{k}\hat{n}\ket{l}|^2 \;,
\end{align}
where $\lambda_{k}$ and $\ket{k}$ are the eigenvalues and corresponding eigenvectors of the state $\rho$.

\subsection{B. Detection noise }

We now consider noise in the measurement stage, as it is inevitable in realistic experiments. We denote $\hat{M}_{\theta}=\cos{\theta} \hat{x} + \sin{\theta} \hat{p}$ an accessible quadrature observable that can be directly measured in continuous-variable platforms. All effective operators in the sets Eqs.~(\ref{SMeq:Mset1}-\ref{SMeq:Mset4}) can be reexpressed in terms of such accessible observables as
\begin{align}
\mathcal{S} \left(\hat{x}\hat{p}\right) &= \hat{M}_{\pi/4}^2 -\frac{1}{2} \hat{M}_0^2 - \frac{1}{2} \hat{M}_{\pi/2}^2, \\
\mathcal{S} \left(\hat{x}^2\hat{p}\right)  &= \frac{ \sqrt{2} }{3} \left( \hat{M}_{\pi/4}^3 + \hat{M}_{3\pi/4}^3 \right) -\frac{1}{3} \hat{M}_{\pi/2}^3 , \\
\mathcal{S} \left( \hat{x}\hat{p}^2 \right)  &= \frac{ \sqrt{2} }{3} \left( \hat{M}_{\pi/4}^3 - \hat{M}_{3\pi/4}^3 \right) -\frac{1}{3} \hat{M}_{0}^3, \\
\mathcal{S} \left(\hat{x}^2\hat{p}^2 \right) &=  \frac{1}{3} \hat{M}_{\pi/4}^4 + \frac{1}{3} \hat{M}_{3\pi/4}^4 -\frac{1}{6} \left( \hat{M}_0^4 + \hat{M}_{\pi/2}^4 \right)  , \\
\mathcal{S} \left(\hat{x}^3\hat{p} \right) & = \frac{2}{\sqrt{3}} \left[ \hat{M}_{\pi/6}^4 - \frac{9}{16} \hat{M}_0^4 -\frac{9}{8} \mathcal{S} \left(\hat{x}^2\hat{p}^2 \right) -\frac{1}{16} \hat{M}_{\pi/2}^4 \right] -\frac{1}{4} \left( \hat{M}_{\pi/4}^4 - \hat{M}_{3\pi/4}^4 \right), \\
\mathcal{S} \left(\hat{x}\hat{p}^3 \right) & = - \frac{2}{\sqrt{3}} \left[ \hat{M}_{\pi/6}^4 - \frac{9}{16} \hat{M}_0^4 -\frac{9}{8} \mathcal{S} \left(\hat{x}^2\hat{p}^2 \right) -\frac{1}{16} \hat{M}_{\pi/2}^4 \right] +  \frac{1}{4} \left( \hat{M}_{\pi/4}^4 - \hat{M}_{3\pi/4}^4 \right).
\end{align}
The complete $k$-th order measurement sets can be written in terms of these accessible observables as
\begin{align}
\mathbf{M}'_{(1)} &= \{ \hat{x}, \hat{p} \} = \{ \hat{M}_{0}, \hat{M}_{\frac{\pi}{2}} \}, \\
\mathbf{M}'_{(2)} &= \{ \mathbf{M}'_{(1)}, \hat{x}^2, \hat{p}^2, \left(\frac{\hat{x}+\hat{p}}{\sqrt{2}} \right)^2 \} = \{ \mathbf{M}'_{(1)}, \hat{M}_{0}^2, \hat{M}_{\frac{\pi}{2}}^2, \hat{M}_{\frac{\pi}{4}}^2 \}, \\
\mathbf{M}'_{(3)} &= \{ \mathbf{M}'_{(2)}, \hat{x}^3, \hat{p}^3, \left(\frac{\hat{x}+\hat{p}}{\sqrt{2}} \right)^3, \left(\frac{-\hat{x}+\hat{p}}{\sqrt{2}} \right)^3 \} = \{ \mathbf{M}'_{(2)}, \hat{M}_{0}^3, \hat{M}_{\frac{\pi}{2}}^3, \hat{M}_{\frac{\pi}{4}}^3, \hat{M}_{\frac{3\pi}{4}}^3 \}, \\
\mathbf{M}'_{(4)} &= \{ \mathbf{M}'_{(3)}, \hat{x}^4, \hat{p}^4, \left(\frac{\hat{x}+\hat{p}}{\sqrt{2}} \right)^4, \left(\frac{-\hat{x}+\hat{p}}{\sqrt{2}} \right)^4, \left(\frac{\sqrt{3}\hat{x}+\hat{p}}{2} \right)^4 \} = \{ \mathbf{M}'_{(3)}, \hat{M}_{0}^4, \hat{M}_{\frac{\pi}{2}}^4, \hat{M}_{\frac{\pi}{4}}^4, \hat{M}_{\frac{3\pi}{4}}^4, \hat{M}_{\frac{\pi}{6}}^4 \}.
\end{align}

We model detection noise as an Gaussian noise on the measurement outcomes. If the ideal measurement of an observable $\hat{M}$ yields the outcome $a$, the actually recorded outcome is $\tilde{a}= a+\Delta M$, where $\Delta M$ is a Gaussian distributed random variable with zero mean value and standard deviation $\sigma$,
\begin{align}
    p(\Delta M)=\frac{1}{\sqrt{2\pi}\sigma }e^{-\frac{\Delta M^2}{2\sigma^2}}.
\end{align}
The moments of Gaussian noise are $\langle (\Delta M)^n \rangle = 0$ for odd $n$, and $\langle (\Delta M)^n \rangle = (n-1)!! \sigma^n$ for even $n$. We write the noisy measurement as
\begin{align}
    \tilde{M} = \hat{M} +\Delta M.
\end{align}
The effect of detection noise on the covariance terms is given by  ($i\neq j$):
\begin{align}
\text{Cov}(\tilde{M}_i, \tilde{M}_i) 
&= \text{Cov} \left( \hat{M}_i, \hat{M}_i \right) + \sigma^2, \\
\text{Cov}(\tilde{M}_i, \tilde{M}_j) 
&= \text{Cov} \left( \hat{M}_i, \hat{M}_j \right) , \\
\text{Cov}(\tilde{M}_i, \tilde{M}_i^2) 
&= \text{Cov} \left( \hat{M}_i, \hat{M}_i^2 \right) + 2 \sigma^2 \langle \hat{M}_i \rangle, \\
\text{Cov}(\tilde{M}_i, \tilde{M}_j^2) 
&= \text{Cov} \left( \hat{M}_i, \hat{M}_j^2 \right), \\
\text{Cov}(\tilde{M}_i, \tilde{M}_i^3) 
&= \text{Cov} \left( \hat{M}_i, \hat{M}_i^3 \right) + 6 \sigma^2 \langle \hat{M}_i^2 \rangle - 3 \sigma^2 \langle \hat{M}_i \rangle^2 + 3\sigma^4, \\
\text{Cov}(\tilde{M}_i, \tilde{M}_j^3) 
&= \text{Cov} \left( \hat{M}_i, \hat{M}_j^3 \right) + 3 \sigma^2 \text{Cov} \left( \hat{M}_i, \hat{M}_j \right), \\
\text{Cov}(\tilde{M}_i, \tilde{M}_i^4) 
&= \text{Cov} \left( \hat{M}_i, \hat{M}_i^4 \right) + 10\sigma^2 \langle \hat{M}_i^3 \rangle -6\sigma^2 \langle \hat{M}_i \rangle \langle \hat{M}_i^2 \rangle +12 \sigma^4 \langle \hat{M}_i \rangle ,\\
\text{Cov}(\tilde{M}_i, \tilde{M}_j^4) 
&= \text{Cov} \left( \hat{M}_i, \hat{M}_j^4 \right) +6 \sigma^2 \text{Cov}\left( \hat{M}_i, \hat{M}_j^2 \right), \\
\text{Cov}( \tilde{M}_i^2, \tilde{M}_i^2) 
&= \text{Cov} \left( \hat{M}_i^2, \hat{M}_i^2 \right) + 4 \sigma^2 \langle \hat{M}_i^2 \rangle +2\sigma^4, \\
\text{Cov}( \tilde{M}_i^2, \tilde{M}_j^2) 
&= \text{Cov} \left( \hat{M}_i^2, \hat{M}_j^2 \right), \\
\text{Cov}(\tilde{M}_i^2, \tilde{M}_i^3) 
&= \text{Cov} \left( \hat{M}_i^2, \hat{M}_i^3 \right) + 9 \sigma^2 \langle \hat{M}_i^3 \rangle -3\sigma^2 \langle \hat{M}_i \rangle \langle \hat{M}_i^2 \rangle +12\sigma^4 \langle \hat{M}_i \rangle , \\
\text{Cov}(\tilde{M}_i^2, \tilde{M}_j^3) 
&= \text{Cov} \left( \hat{M}_i^2, \hat{M}_j^3 \right) + 3 \sigma^2 \text{Cov} \left( \hat{M}_i^2,  \hat{M}_j \right), \\
\text{Cov}(\tilde{M}_i^2, \tilde{M}_i^4) 
&= \text{Cov} \left( \hat{M}_i^2, \hat{M}_i^4 \right) +  14 \sigma^2 \langle \hat{M}_i^4 \rangle - 6\sigma^2 \langle \hat{M}_i^2 \rangle^2 + 36 \sigma^4 \langle \hat{M}_i^2 \rangle +12 \sigma^6, \\
\text{Cov}(\tilde{M}_i^2, \tilde{M}_j^4) 
&= \text{Cov} \left( \hat{M}_i^2, \hat{M}_j^4 \right) + 6 \sigma^2 \text{Cov} \left( \hat{M}_i^2, \hat{M}_j^2 \right), \\
\text{Cov}(\tilde{M}_i^3, \tilde{M}_i^3) 
&= \text{Cov} \left( \hat{M}_i^3, \hat{M}_i^3 \right) + 15 \sigma^2 \langle \hat{M}_i^4 \rangle -6\sigma^2 \langle \hat{M}_i \rangle \langle \hat{M}_i^3 \rangle +45 \sigma^4 \langle \hat{M}_i^2 \rangle -9 \sigma^4 \langle \hat{M}_i \rangle^2 +15\sigma^6 , \\
\text{Cov}(\tilde{M}_i^3, \tilde{M}_j^3) 
&= \text{Cov} \left( \hat{M}_i^3, \hat{M}_j^3 \right) + 3 \sigma^2 \text{Cov} \left( \hat{M}_i^3,  \hat{M}_j \right) + 3\sigma^2 \text{Cov} \left(  \hat{M}_i, \hat{M}_j^3 \right) + 9 \sigma^4 \text{Cov} \left( \hat{M}_i,  \hat{M}_j \right), \\
\text{Cov}(\tilde{M}_i^3, \tilde{M}_i^4) 
&= \text{Cov} \left( \hat{M}_i^3, \hat{M}_i^4 \right) + 21\sigma^2 \langle \hat{M}_i^5 \rangle -6\sigma^2 \langle \hat{M}_i^2 \rangle \langle \hat{M}_i^3 \rangle -3\sigma^2 \langle \hat{M}_i \rangle \langle \hat{M}_i^4\rangle \nonumber \\
&\quad -18 \sigma^4 \langle \hat{M}_i \rangle \langle \hat{M}_i^2 \rangle +102 \sigma^4 \langle \hat{M}_i^3 \rangle + 96 \sigma^6 \langle \hat{M}_i \rangle, \\
\text{Cov}(\tilde{M}_i^3, \tilde{M}_j^4) 
&= \text{Cov} \left( \hat{M}_i^3, \hat{M}_j^4 \right) + 6 \sigma^2 \text{Cov} \left( \hat{M}_i^3, \hat{M}_j^2 \right) + 3 \sigma^2 \text{Cov} \left( \hat{M}_i, \hat{M}_j^4 \right) + 18 \sigma^4 \text{Cov} \left( \hat{M}_i, \hat{M}_j^2 \right), \\
\text{Cov}(\tilde{M}_i^4, \tilde{M}_i^4) 
&= \text{Cov} \left( \hat{M}_i^4, \hat{M}_i^4 \right) + 28 \sigma^2 \langle \hat{M}_i^6\rangle -12\sigma^2 \langle \hat{M}_i^2 \rangle \langle \hat{M}_i^4 \rangle + 204 \sigma^4 \langle \hat{M}_i^4 \rangle -36 \sigma^4 \langle \hat{M}_i^2 \rangle^2 +384 \sigma^6 \langle \hat{M}_i^2 \rangle +96 \sigma^8 , \\
\text{Cov}(\tilde{M}_i^4, \tilde{M}_j^4) 
&= \text{Cov} \left( \hat{M}_i^4, \hat{M}_j^4 \right) +6\sigma^2 \text{Cov} \left( \hat{M}_i^4, \hat{M}_j^2 \right) +6\sigma^2 \text{Cov} \left( \hat{M}_i^2, \hat{M}_j^4 \right) +36 \sigma^4 \text{Cov} \left( \hat{M}_i^2, \hat{M}_j^2 \right)
\end{align}

Analogously, its effect on the commutator terms read
\begin{align}
\langle[ \hat{n}, \tilde{M}_i ] \rangle
&= \langle [ \hat{n}, \hat{M}_i +\Delta \hat{M}_i ] \rangle
= \langle [ \hat{n}, \hat{M}_i ] \rangle ,\\
\langle [ \hat{n}, \tilde{M}_i^2 ] \rangle &= \langle [ \hat{n}, \hat{M}_i^2 +2\Delta \hat{M}_i \hat{M}_i + \Delta M_i^2 ] \rangle = \langle [ \hat{n}, \hat{M}_i^2 ] \rangle, \\
\langle [ \hat{n}, \tilde{M}_i^3 ] \rangle &= \langle [ \hat{n}, \hat{M}_i^3 +3\Delta \hat{M}_i \hat{M}_i^2 +3\Delta \hat{M}_i^2 \hat{M}_i + \Delta \hat{M}_i^3 ] \rangle = \langle [ \hat{n}, \hat{M}_i^3 ] \rangle  + 3\sigma^2  \langle [\hat{n},  \hat{M}_i ] \rangle, \\
\langle [ \hat{n}, \tilde{M}_i^4 ] \rangle &= \langle [ \hat{n}, \hat{M}_i^4 +4\Delta \hat{M}_i \hat{M}_i^3 +6\Delta \hat{M}_i^2 \hat{M}_i^2  +4\Delta \hat{M}_i^3 \hat{M}_i + \Delta \hat{M}_i^4 ] \rangle = \langle [ \hat{n}, \hat{M}_i^4 ] \rangle +  6\sigma^2 \langle [\hat{n},  \hat{M}_i^2 ] \rangle.
\end{align}

Based on these covariance matrix and commutator matrix expressed in terms of the noisy measurements $\tilde{M}_i^k$, we obtain the corresponding nonlinear squeezing coefficients in the presence of detection noise, as shown in Fig.~\ref{Fig3}(c) of the main text.

\section{IV.\quad Practical protocols to generate cubic phase states }

Cubic phase states are central to universal CV quantum computation, yet their preparation is challenging due to the lack of natural implementations of the gate $\hat{C}=e^{ir\hat{x}^3}$. Consequently, several approximate schemes have been developed, including protocols based on non-Gaussian measurements~\cite{MarshallPRA2015,MarekPRA2011,MiyataPRA2016,JingarXiv2025} and higher-order interactions~\cite{YanagimotoPRL2020,ZhengPRXQ2021}. Weak cubic states have been demonstrated in optical platforms~\cite{YukawaPRA2013}, and exact cubic phase states have recently been realized deterministically in superconducting microwave circuits~\cite{ErikssonNC2024}. In this section, we analyze the metrological performance of states produced by these preparation methods.

\subsection{A. Repeat-until-success protocol based on sequential photon subtraction }

One practical method to generate cubic phase state is to implement non-Gaussian measurements, such as photon subtraction, together with Gaussian operations. Here we analyze the repeat-until-success protocol in Ref.~\cite{MarshallPRA2015} as a representative example. In this scheme, the cubic phase gate $\hat{C}=e^{i r \hat{x}^3}$ is approximated by the decomposition
\begin{align}
\hat{C}_N(r) = \left( 1+ i \frac{r}{N} \hat{x}^3 \right)^N,
\end{align}
where $N$ is the number of iterations. In each iteration, the operator $\left( 1+ i \frac{r}{N} \hat{x}^3 \right)$ is realized via a combination of photon subtraction and Gaussian operations. In the limit of large $N$, these operations converge to the ideal cubic gate. Starting from the squeezed vacuum state $e^{ -\frac{1}{2} s\left( \hat{a}^2 - \hat{a}^{\dagger 2} \right)} |0\rangle$, repeated application of the approximate cubic operation $N$ times produces
\begin{align}\label{SMeq:RUSstate}
|\psi(r,s,N) \rangle &= \frac{1}{\sqrt{Z_N}} \left( 1+ i \frac{r}{N} \hat{x}^3 \right) e^{ -\frac{1}{2} s\left( \hat{a}^2 - \hat{a}^{\dagger 2} \right)} |0\rangle,
\end{align}
where $Z_N$ is a normalization parameter. The corresponding wave function reads
\begin{align}
\psi_N(x) = \frac{1}{\sqrt{Z_N}} \left( 1+ i \frac{r}{N} x^3 \right)^N  \frac{1}{ (\pi e^{2s})^{1/4} } e^{ - \frac{x^2}{2e^{2s}}}.
\end{align}
We derive analytical expressions for the normalization values $Z_N$ for $N=\{1,\cdots,5\}$,
\begin{align}
Z_1 &= 1 + \frac{15}{8}  e^{6s} r^2, \\
Z_2 &= 1+ \frac{15}{16} e^{6s} r^2 + \frac{10395 r^4 e^{12s} }{1024}, \\
Z_3 &= 1+ \frac{5}{8} e^{6s} r^2 + \frac{385}{64} e^{12s} r^4 + \frac{425425 }{4608} e^{18s} r^6, \\
Z_4 &= 1 + \frac{15}{32} e^{6s} r^2 + \frac{31185}{8192} e^{12s}r^4 + \frac{ 34459425 }{ 524288 } e^{18s} r^6 + \frac{ 316234143225  }{ 268435456 } e^{24 s} r^8, \\
Z_5 &= 1 + \frac{3}{8} e^{6s} r^2 + \frac{2079}{800} e^{12s} r^4 + \frac{ 1378377  }{ 32000 } e^{18s} r^6 + \frac{ 12649365729  }{ 12800000 } e^{24s} r^8 + \frac{ 9904453365807  }{ 512000000 } e^{30s} r^{10}  .
\end{align}
From these wave functions, we calculate the expectation values $\langle \hat{n} \rangle$ 
\begin{align}
\langle \hat{n} \rangle_{|\psi(r,s,N) \rangle} &= \int_{-\infty}^{\infty} \text{d}x \psi_N^*(x) \frac{1}{2}\left( x^2 - \partial_x^2 -1 \right) \psi_N^*(x) 
\end{align}
which yields
\begin{align}
\langle \hat{n} \rangle_{|\psi(r,s,N=1) \rangle}&= \frac{ \sinh^2(s) + \frac{3}{16} e^{6s} r^2 \left( -5 +23\cosh{(2s)} + 12\sinh{(2s)} \right) }{1+ \frac{15}{8}r^2 e^{6s}  }, \\
\langle \hat{n} \rangle_{|\psi(r,s,N=2) \rangle} &= \frac{1}{4}\left( -2 +e^{-2s} +13e^{2s} + \frac{ 12e^{2s}\left(-1024 +3e^{2s}r^2 \left(224-160e^{4s} +315 e^{6s} r^2 \right) \right) }{ 1024 +15e^{6s} r^2 \left( 64 +693e^{6s}r^2 \right) } \right), \\
\langle \hat{n} \rangle_{|\psi(r,s,N=3) \rangle} &= \frac{1}{4}\left[ -2 +e^{-2s} +19 e^{2s} + \frac{ 18 e^{2s} \left[ -4608 +e^{2s} r^2 \left( 2496 +5e^{4s} \left( -384 +7e^{2s} r^2 (336 -264 e^{4s} +715 e^{6s} r^2 ) \right) \right) \right] }{  4608 +5e^{6s}r^2 \left( 576 + 77e^{6s} r^2 (72 +1105 e^{6s} r^2 ) \right) }  \right], \\
\langle \hat{n} \rangle_{|\psi(r,s,N=4) \rangle} &= \frac{1}{4} \Big[ -2 +e^{-2s} +25 e^{2s} + \frac{ 1 }{ 268435456 + 15e^{6s} r^2 \left[ 8388608 + 2079 e^{6s} r^2 \left( 32768 +1105 e^{6s} r^2 \left( 512+9177 e^{6s} r^2 \right) \right) \right] } \nonumber \\
& \quad \times 24 e^{2s} \Big[ -268435456 + 3 e^{2s} r^2 \left( 39845888 -31457280 e^{4s} \right. \nonumber \\
& \quad \left. + 945 e^{6s} r^2 \left( 212992 -180224 e^{4s} +715 e^{6s} r^2 \left( 2688 -2176 e^{4s} + 6783 e^{6s} r^2 \right) \right) \right) \Big]  \Big] ,\\
\langle \hat{n} \rangle_{|\psi(r,s,N=5) \rangle} &= \frac{1}{4} \Big[ -2+ e^{-2s} +31 e^{2s} \notag\\&\quad+\frac{1}{ 512000000 + 3e^{6s} r^2 \left( 64000000 + 693 e^{6s} r^2 \left( 640000 +663 e^{6s} r^2 \left(  16000 +9177 e^{6s} r^2 (40 +783 e^{6s} r^2) \right) \right) \right)  } \nonumber \\
& \quad \times 30 e^{2s} \Big[ -512000000 +3e^{2s} r^2 \Big[  64000000 -51200000e^{4s} + 63 e^{6s} r^2 \Big[  4864000 -4224000 e^{4s} \nonumber \\
& \quad +429 e^{6s} r^2 \left( 124800 +17e^{4s}\left( -6400+399e^{2s}r^2(224-184e^{4s} +621 e^{6s}r^2 ) \right) \right)
\Big]
\Big]
\Big]
\Big].
\end{align}
The second moments $\langle \hat{n}^2 \rangle_{|\psi(r,s,N) \rangle}$ can be derived analogously,
\begin{align}
\langle \hat{n}^2 \rangle _{|\psi(r,s,N) \rangle}&= \int_{-\infty}^{\infty} \text{d}x \psi_N^*(x) \frac{1}{4}\left( x^2 - \partial_x^2 -1 \right)^2 \psi_N^*(x),
\end{align}
which leads to the analytical expressions 
\begin{align}
\langle \hat{n}^2 \rangle_{|\psi(r,s,N=1) \rangle} &=  \frac{ e^{-4s} \left[ 24 -32e^{2s} +16e^{4s} +e^{6s} \left( -32 +225 r^2 +3 e^{2s} \left( 8 + r^2 \left( -44 +5e^{2s} \left( 26 -28 e^{2s} +63 e^{4s} \right) \right) \right) \right)  \right]  }{ 16 \left( 8 +15e^{6s} r^2 \right)}, \\
\langle \hat{n}^2\rangle_{|\psi(r,s,N=2) \rangle} &= \frac{1}{16} \Big[ 50 + 3e^{-4s} - 4e^{-2s} -52 e^{2s} +195 e^{4s} \nonumber \\
& \quad + \frac{ 48 \left( -1024 +960 e^{6s} r^2 -32 e^{4s} (128 +21 r^2) + 16 e^{2s} (64+45 r^2) -15 e^{10s} r^2 (176 +63 r^2) +15 e^{8s} r^2 (32 +147 r^2) \right)}{ 1024 +15 e^{6s} r^2(64 +693 e^{6s} r^2) } \Big] ,\\
\langle \hat{n}^2 \rangle_{|\psi(r,s,N=3) \rangle} &= \frac{1}{16} \Big[ 74 +3 e^{-4s} -4e^{-2s} -76 e^{2s} +399 e^{4s} + \frac{ 1 }{ 4608 +5e^{6s} r^2 \left( 576 +77e^{6s} r^2 (72 +1105 e^{6s} r^2) \right) } \nonumber \\
& \quad \times 36 \Big[ -9216 +7680 e^{6s} r^2 +92400 e^{12s} r^4 -3360 e^{10s} r^2 (8+7 r^2) -384 e^{4s} (132+13 r^2) \nonumber \\
& \quad +192 e^{2s} (48 +25 r^2) -770 e^{16s }r^4 (204 +65 r^2) +385 e^{14s} r^4 (48 +299 r^2) +3840 e^{8s} (r^2 +14 r^4) \Big] \Big], \\
\langle \hat{n}^2 \rangle_{|\psi(r,s,N=4) \rangle} &= \frac{1}{ 16 \left( 268435456 + 15 e^{6s} r^2 \left( 8388608 +2079 e^{6s} r^2 \left( 32768 +1105 e^{6s} r^2 (512 +9177 e^{6s} r^2 ) \right) \right) \right) } \nonumber \\
& \quad \times e^{-4s} \Big[ 805306368 -1073741824 e^{2s} + 536870912 e^{4s} + 30450647040 e^{10s} r^2 + 345392087 040 e^{16s} r^4 \nonumber \\
& \quad + 5116535424000 e^{22s} r^6 +30990946036050 e^{28s} r^8 - 31623414322500 e^{30s} r^8 +213458046676875 e^{32s} r^8 \nonumber \\
& \quad -54997242300 e^{26s} r^6 (-128 +47 r^2) -100663296 e^{8s} (-8+119r^2) -4151347200 e^{20s} r^4 (-48 +143r^2) \nonumber \\
& \quad - 123863040 e^{14s} r^2 (-64+501 r^2) + 8388608 e^{6s} (-128 +1305 r^2) + 654729075 e^{24s} r^6 (-2048 + 6057r^2) \nonumber \\
& \quad +207567360 e^{18s} r^4 (-256 +6627 r^2) + 3440640 e^{12s} r^2 (-1024 +54891 r^2)
\Big] ,
\end{align}
\begin{align}
\langle \hat{n}^2 \rangle_{|\psi(r,s,N=5) \rangle} &= \frac{1}{ 16\left( 512000000 +3e^{6s} r^2 \left( 64000000 + 693 e^{6s} r^2 \left( 640000 +663 e^{6s} r^2 \left(16000 + 9177 e^{6s} r^2 (40+783 e^{6s} r^2 ) \right) \right) \right) \right) } \nonumber \\
& \quad \times e^{-4s} \Big[ 1536000000 - 2048000000 e^{2s} + 1024000000 e^{4s} +  60288000000 e^{10s} r^2 + 641329920000 e^{16s} r^4 \nonumber \\
& \quad + 11159340192000 e^{22s} r^6 + 
 195306206855760 e^{28s} r^8 + 
 1208343310628454 e^{34s} r^{10} - 
 1228152217360068 e^{36s} r^{10} \nonumber \\
& \quad +10132255793220561 e^{38s}r^{10} - 
 768000000 e^{8s} (-2+31r^2) - 
 1366131498732 e^{32s} r^8 (-250 +59 r^2) \nonumber \\
& \quad -87995587680 e^{26s}r^6 (-100 +191 r^2) - 483840000 e^{14s} r^2 (-25 +239 r^2) - 5189184000 e^{20s} r^4 (-50 +251 r^2) \nonumber \\
 & \quad +64000000 e^{6s} (-32 +333r^2) + 
 12649365729 e^{30s} r^8 (-4000+9729r^2) + 
 432432000 e^{18s} r^4 (-160 +10413 r^2) \nonumber\\
 & \quad +13440000 e^{12s} r^2 (-400+34569r^2) + 1047566520 e^{24s} r^6 (-1600+38601 r^2)
\Big].
\end{align}
Since the state in Eq.~\eqref{SMeq:RUSstate} is pure, its QFI is determined by the variance, $F_Q[\psi(r,s,N),\hat{n}] = 4\text{Var}[\psi(r,s,N),\hat{n}] = 4\left( \langle \hat{n}^2 \rangle_{|\psi(r,s,N) \rangle} - \langle \hat{n} \rangle_{|\psi(r,s,N) \rangle}^2 \right)$.

In Fig.~\ref{FigSM_RepeatN}, we scan the squeezing strength $s=[0,2]$ and the cubicity $r=[0,0.4]$, and compute $F_Q/n$ for the states generated by $N$ iterations as a function of population $n$. The maximum $F_Q^{\text{max}}/n$ for each $N$ is extracted and plotted in Fig.~\ref{Fig4} of the main text. It is observed that the maximum sensitivity $F^{\text{max}}_Q/n$ of the approximative cubic phase states surpasses that of the squeezing vacuum states. As $N$ increases, $F^{\text{max}}_Q/n$ of the approximative states grows and approaches the sensitivity of ideal cubic phase states.

\begin{figure}[t]
    \centering
    \includegraphics[width=120mm]{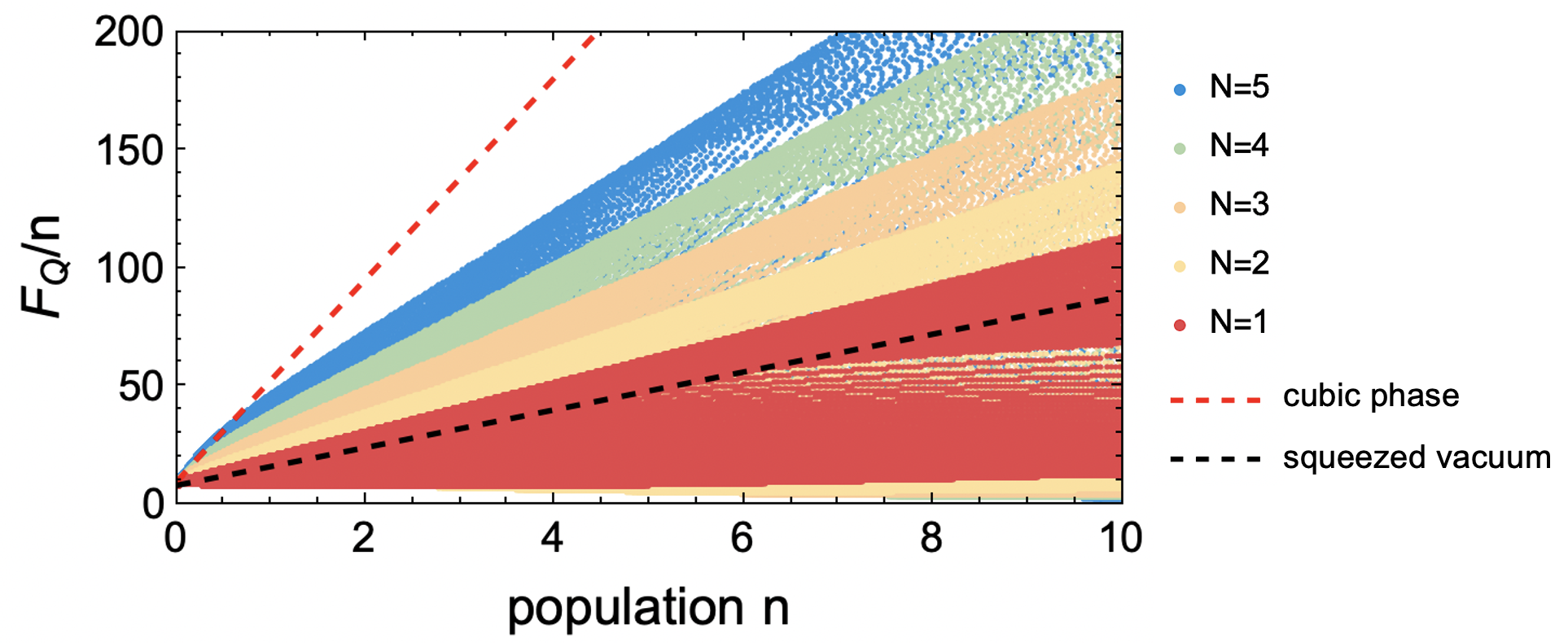}
    \caption{\textbf{Sensitivity of approximate cubic phase states generated from the repeat-until-success protocol.} we scan the squeezing strength $s=[0,2]$ and the cubicity $r=[0,0.4]$, and plot $F_Q/n$ as a function of population for different number of iterations $N$. }
    \label{FigSM_RepeatN}
\end{figure}

\subsection{B. Kerr-based Hamiltonian with Gaussian channels }
Another approach to realizing the cubic phase states is to use nonlinear interactions. A paradigmatic model of such nonlinearity the Kerr interaction $e^{iK\hat{a}^{\dagger 2} \hat{a}^2 }$, where $K$ is the Kerr parameter. Since the Kerr unitary commutes with the number operator, 
\begin{align}
\left[ e^{iK\hat{a}^{\dagger 2} \hat{a}^2 }, \hat{n} \right] &= 0,
\end{align}
it fails to enhance sensitivity in rotation sensing scenarios. We also consider an experimentally relevant Hamiltonian in Refs.~\cite{GuoPRA2024,VenkatramanPNAS2024}, $\hat{H} = \Delta \hat{a}^\dagger \hat{a} + s \left( \hat{a}^{\dagger 2} + \hat{a}^2 \right) - K\hat{a}^{\dagger 2} \hat{a}^2$, but the resulting sensitivity $F_Q/n$ still does not surpass that of squeezed vacuum states.

A protocol of approximating cubic phase state based on Kerr interaction is proposed in Ref.~\cite{YanagimotoPRL2020}. In this scheme, a Kerr-driven evolution is sandwiched between two Gaussian channels, yielding an effective cubic phase gate 
\begin{align}\label{SMeq:CKerrEff}
\hat{C}_{\text{Kerr}}^{\text{eff}} &= \hat{S}^\dagger (\text{log} \lambda ) \hat{D}^\dagger (\alpha) e^{-i \hat{H}_{\text{Kerr}} \tau } \hat{D}(\alpha) \hat{S}(\text{log} \lambda).
\end{align}
Here, $\hat{S}(\text{log} \lambda )$ is a squeezing operator with in-phase quadrature power gain $\lambda^2$ ($\lambda>1$), and $\hat{D}^\dagger (\alpha)$ is a displacement operator. The Kerr Hamiltonian $\hat{H}_{\text{Kerr}}$ reads
\begin{align}
\hat{H}_{\text{Kerr}} &= - \frac{K}{2} \hat{a}^{\dagger 2} \hat{a}^2 + \Delta \hat{a}^\dagger \hat{a} + \beta \left( \hat{a} +\hat{a}^\dagger \right),
\end{align}
where $\Delta$ denotes the detuning and $\beta$ is the driving amplitude. By choosing $\Delta = 3K \alpha^2 -K$ and $\beta= -2K\alpha^3$, undesired terms $\hat{x}^2$ and $\hat{x}$ are removed, and the effective Hamiltonian in Eq.~\eqref{SMeq:CKerrEff} becomes
\begin{align}
\hat{H}_{\text{Kerr}}^{\text{eff}} &= - \frac{K}{8} \left( \lambda^4 \hat{x}^4 + \hat{p}\hat{x}^2 \hat{p} + \hat{x} \hat{p}^2 \hat{x} + \lambda^{-4} \hat{p}^4 \right) - \frac{1}{\sqrt{2}} K\lambda^3 \alpha \hat{x}^3 - \frac{1}{\sqrt{2}} K \lambda^{-1} \alpha \hat{p} \hat{x} \hat{p} + K \lambda^{-2}  \alpha^2 \hat{p}^2.
\end{align}
If one assumes $\alpha\sim \lambda^3$, the Kerr-based cubic phase Hamiltonian can be written as
\begin{align}
\hat{H}_{\text{Kerr}}^{\text{eff}} &= - \frac{ K\lambda^3 \alpha }{ \sqrt{2} } \left( \hat{x}^3 + \frac{ \lambda \alpha^{-1} }{ 4\sqrt{2} } \hat{x}^4 -\sqrt{2} \lambda^{-5} \alpha \hat{p}^2 + \mathcal{O}\left( \lambda^{-4} \right) \right).
\end{align}
Setting the evolution time to $\tau = \sqrt{2} r/(K\alpha \lambda^3) $, where $r$ is the target cubicity, the approximate cubic phase gate is obtained as
\begin{align}\label{SMeq:CKerrCubic}
\hat{C}_{\text{Kerr}}^{\text{cubic}} &= \text{exp}\left( -i \tau \hat{H}_{\text{Kerr}}^{\text{eff}} \right) \nonumber \\
&= \exp{\left[ i r  \left( \hat{x}^3 +\frac{\lambda \alpha^{-1} }{ 4\sqrt{2} } \hat{x}^4 -\sqrt{2}\lambda^{-5} \alpha \hat{p}^2 +\mathcal{O}(\lambda^{-4})  \right) \right]}.
\end{align}
If the parameters determining the Gaussian channels are chosen such that the prefactors $\lambda \alpha^{-2}$ and $\lambda^{-5}\alpha$ are negligible, the gate in Eq.~\eqref{SMeq:CKerrCubic} approaches the ideal cubic phase gate. An approximate cubic phase state is then generated by applying  $\hat{C}_{\text{Kerr}}^{\text{cubic}}$ to a squeezed vacuum state, $\hat{C}_{\text{Kerr}}^{\text{cubic}} e^{-\frac{1}{2} s\left( \hat{a}^2 -\hat{a}^{\dagger 2} \right) } |0\rangle $. Following this protocol, we set the parameters in Gaussian channels to squeezing $\lambda=[2,5]$ and displacement amplitude $\alpha=\lambda^3$, and then scan over the cubicity $r$ and squeezing strength $s$. Fig.~\ref{FigSM_KerrPoints} shows that the metrological sensitivity of the approximate cubic phase states approaches that of the ideal cubic phase states.

\begin{figure}[h]
    \centering
    \includegraphics[width=120mm]{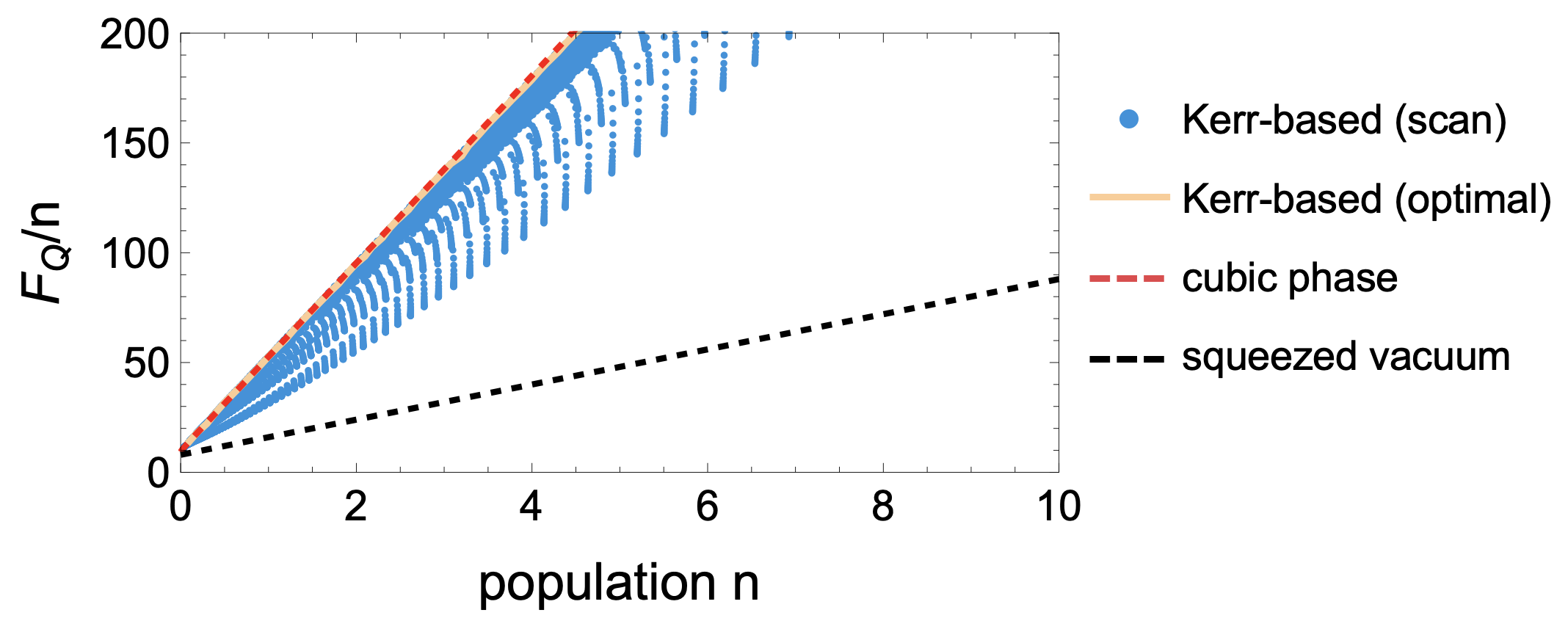}
    \caption{\textbf{Sensitivity of approximate cubic phase states generated from the Kerr-based protocol.} we restrict the parameters in Gaussian channels to $\lambda=[2,5]$ and $\alpha=\lambda^3$, and scan over the cubicity $r$ and squeezing strength $s$ to obtain the corresponding $F_Q/n$ as a function of population $n$ (blue points). The maximum sensitivity (yellow line) coincides with that of ideal cubic phase states (red dashed line). }
    \label{FigSM_KerrPoints}
\end{figure}

\subsection{C. Trisqueezing interaction }
The trisqueezing interaction
\begin{align}
\hat{U}_T &= e^{i \left( t^* \hat{a}^3 +t\hat{a}^{\dagger 3} \right) },
\end{align}
is another typical nonlinearity that can be used to prepared non-Gaussian states, where $t$ denotes the triplicity. This interaction can be realized by three-photon spontaneous parametric down-conversion, that has been experimentally implemented in a superconducting parameter cavity~\cite{ChangPRX2020}. This trisqueezing interaction can be used to prepare cubic phase states: Ref.~\cite{ZhengPRXQ2021} proposed a protocol to approximate the cubic phase gate by applying Gaussian operations on a trisqueezing state, and Ref.~\cite{ErikssonNC2024} experimentally generated a cubic phase states in superconducting microwave circuits using a universal set of generalized squeezing operations with the trisqueezing interaction involved.

In addition to their utility for preparing cubic phase states, we find that trisqueezed states themselves have a high metrological potential in rotation sensing tasks. However, these states are not analytically tractable and, as the triplicity grows ($t \gtrsim 0.1$), the dimension required to fully characterize the state is growing rapidly. Below, we present partial results for population and sensitivity with different truncation dimensions in Fig.~\ref{FigSM_trisqueezed}. Even though these results have not converged in the intermediate region of $t\simeq 1$ in Fig.~\ref{FigSM_trisqueezed}, the population returns to small values around $1.6\leq t\leq 1.9$ indicating an almost periodic behavior in their evolution---see Ref.~\cite{SahelNJP2025} for a more detailed discussion. In Fig.~\ref{FigSM_trisqueezed}(b), we compare the ultimate sensitivity $F_Q/n$ of the trisqueezed states $\hat{U}_T|0\rangle$ to the sensitivity of cubic phase states and squeezed vacuum states. Even though the numerical data is not fully converged, it indicates that the sensitivity of trisqueezed states surpass that of cubic phase states for small populations. However, periodic dynamics~\cite{SahelNJP2025} of the trisqueezed states limit the maximum population $n$ that can be reached from increasing the triplicity $t$. Consequently, cubic phase states and any family of states whose sensitivity grows with $n$ will outperform the trisqueezed states for large populations.

\begin{figure}[h]
    \centering
    \includegraphics[width=\textwidth]{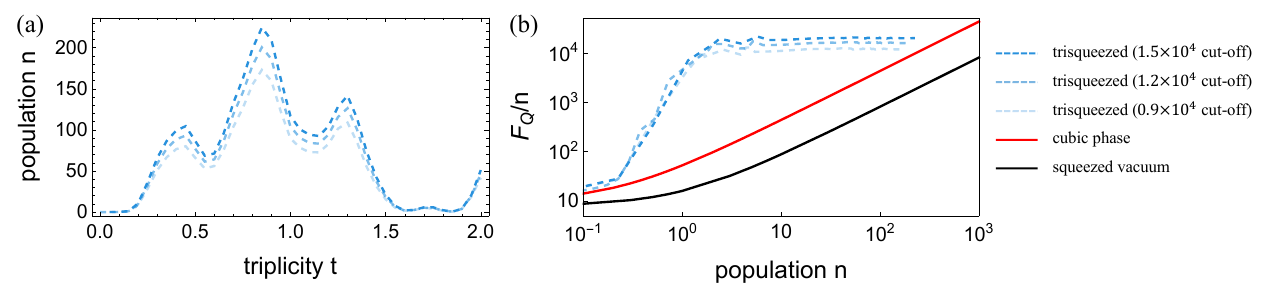}
    \caption{\textbf{Oscillatory dynamics and metrological sensitivity of trisqueezed states.} (a) The population $n$ follows oscillatory dynamics as the triplicity $r$ increases. (b) We compare the sensitivity of trisqueezed states for different cut-off number to that of cubic phase states and squeezed vacuum states.  }
    \label{FigSM_trisqueezed}
\end{figure}

\end{widetext}

\end{document}